%% file: MAIN.tex
\documentclass[final,3p,times,twocolumn]{elsarticle}

\usepackage{graphicx}
\usepackage{amssymb}

\usepackage{lineno}

\biboptions{sort&compress}
\usepackage{algorithmic}
\usepackage[ruled,vlined,commentsnumbered]{algorithm2e}
\usepackage{multirow}
\usepackage{array}
\usepackage{mdwmath}
\usepackage{mdwtab}
\usepackage{eqparbox}
\usepackage{url}
\usepackage[cmex10]{amsmath}
\usepackage{xcolor}

\journal{Artificial Intelligence in Medicine}

\begin{document}

\begin{frontmatter}


\title{Circumpapillary OCT-Focused Hybrid Learning for Glaucoma Grading Using Tailored Prototypical Neural Networks}

\ead{jogarpa7@i3b.upv.es}



\author{Gabriel Garc\'ia\textsuperscript{1}, Roc\'io del Amor\textsuperscript{1}, Adri\'an Colomer\textsuperscript{1}, Rafael Verd\'u-Monedero\textsuperscript{2}, Juan Morales-S\'anchez\textsuperscript{2} and Valery Naranjo\textsuperscript{1}}

\address{\textsuperscript{1}Instituto de Investigación e Innovación en Bioingeniería, Universitat Politècnica de València, 46022, Valencia, Spain}
\address{\textsuperscript{2}Departamento de Tecnologías de la Información y las Comunicaciones, Universidad Politécnica de Cartagena, 30202, Cartagena, Spain}

\input{0_Abstract}

\end{frontmatter}


\input{1_Introduction}

\input{2_0_Related_work}

\input{3_0_Methods}

\input{4_0_Ablation_experiments}

\input{5_Prediction_results}

\input{6_Discussion}

\input{7_Conclusion}



\section*{Funding}

This work has been partially funded by the following projects: [H2020-ICT-2016-2017, 732613], DPI2016-77869-C2-1-R, PROMETEO/2019/109, AES2017-PI17/00771, AES2017-PI17/00821 and 20901/PI/18. The work of Gabriel Garc\'{i}a has been supported by the Spanish State Research Agency PTA2017-14610-I.







\bibliographystyle{elsarticle-num}
\bibliography{refs.bib}







\end{document}

%% file: 0_Abstract.tex
\begin{abstract}

Glaucoma is one of the leading causes of blindness worldwide and Optical Coherence Tomography (OCT) is the quintessential imaging technique for its detection. Unlike most of the state-of-the-art studies focused on glaucoma detection, in this paper, we propose, for the first time, a novel framework for glaucoma grading using raw circumpapillary B-scans. In particular, we set out a new OCT-based hybrid network which combines hand-driven and deep learning algorithms. An OCT-specific descriptor is proposed to extract hand-crafted features related to the retinal nerve fibre layer (RNFL). In parallel, an innovative CNN is developed using skip-connections to include tailored residual and attention modules to refine the automatic features of the latent space. The proposed architecture is used as a backbone to conduct a novel few-shot learning based on static and dynamic prototypical networks. The $k-$shot paradigm is redefined giving rise to a supervised end-to-end system which provides substantial improvements discriminating between healthy, early and advanced glaucoma samples. The training and evaluation processes of the dynamic prototypical network are addressed from two fused databases acquired via Heidelberg Spectralis system. Validation and testing results reach a categorical accuracy of 0.9459 and 0.8788 for glaucoma grading, respectively. Besides, the high performance reported by the proposed model for glaucoma detection deserves a special mention. The findings from the class activation maps are directly in line with the clinicians' opinion since the heatmaps pointed out the RNFL as the most relevant structure for glaucoma diagnosis.

\end{abstract}

\begin{keyword}
Glaucoma Grading \sep Prototypical Neural Networks \sep Circumpapillary \sep Hybrid Learning \sep Retinal Nerve Fiber Layer \sep Optical Coherence Tomography
\end{keyword}

%% file: 1_Introduction.tex
\section{Introduction}\label{sec:01_Introduction}

Glaucoma is a chronic and progressive disease that affects the optic nerve head (ONH) of the retina causing several structural changes and functional damage \cite{weinreb2004}. Nowadays, this optic neuropathy has become the leading cause of blindness worldwide, according to \cite{jonas2018}. Recent studies suggest that the impact of this disease will continue to rise, affecting 111.8 million people in 2040 \cite{tham2014}. Therefore, early diagnosis of glaucoma could be essential for timely treatment in order to prevent irreversible vision loss \cite{jonas2018}.

\begin{figure*}[h]
\begin{center}
\includegraphics[width=16cm]{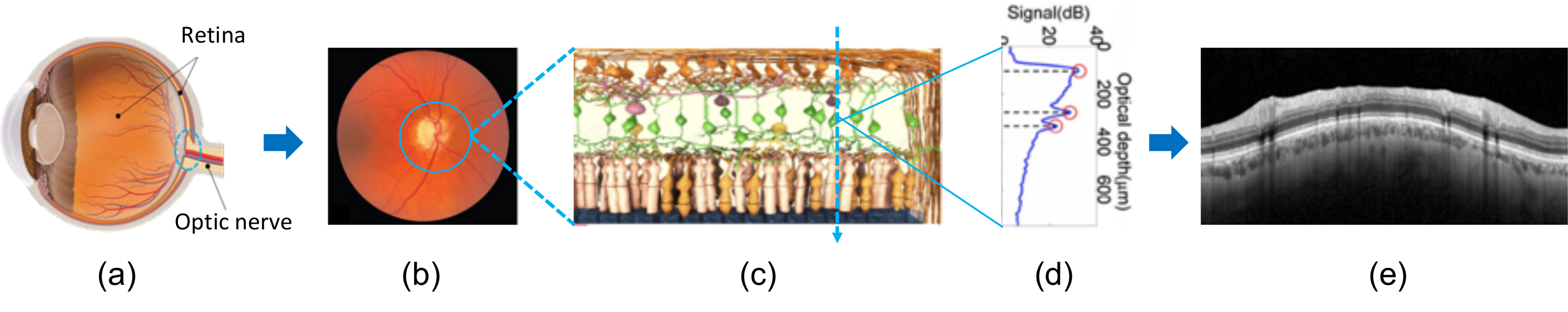}\\
\end{center}
\caption{(a) Eyeball with interesting regions. (b) Fundus image focused on the optic nerve head (ONH). (c) Arrangement of the retinal fiber layers. (d) A-scan corresponding to the depth of retinal cells at a specific point. (e) B-scan representing the fiber layers of the retina with different gray-intensity levels.}
\label{OCT_ac}
\end{figure*}

Currently, there is no single accurate test to certify glaucoma, the diagnostic procedure includes several time-consuming tests such as pachymetry, tonometry and visual field tests, as well as the examination of different kinds of interpretable retinal images. Specifically, techniques based on image analysis like fundus photography and optical coherence tomography (OCT) have become very important in the context of glaucoma detection. Fundus image is a great cost-effectiveness technique which has reported promising results in the diagnosis of several eye-focused diseases, e.g. diabetic retinopathy \cite{floriano2019, gulshan2016} and age-related macular degeneration \cite{burlina2017, grassmann2018}. However, OCT imaging modality \cite{huang1991} is the quintessential technique for glaucomatous damage evaluation \cite{bussel2014} since it allows quantifying glaucoma-specific regions such as retinal nerve fibre layer (RNFL) and ganglion cell inner plexiform layer (GCIPL), which are useful biomarkers for the progression of this disease \cite{medeiros2009}. Additionally, glaucoma is evident in the deterioration of the cell layers around the optic disc, whose information could be exploited by the OCT imaging modality since it focuses on the depth axis of the retina to identify structural changes, unlike the 2D projection of the fundus image (see Fig. \ref{OCT_ac}). 

Note that, although fundus image modality is cheaper than OCT, it is colour-dependent on the training data set and its interpretation remains subjective \cite{lichter1976, varma1992}. The contrary, OCT is a non-contact and non-invasive technique that provides objective information about the optic nerve head and RNFL structures \cite{jaffe2004}. It is important to remark that glaucoma detection entails a subjective examination from different experts, whose mismatch ratio is usually high \cite{national2017}. Consequently, many state-of-the-art studies developed different machine learning algorithms intended to detect glaucoma via fundus image and OCT samples. 

%% file: 2_0_Related_work.tex
\section{Related work}\label{sec:2_Related_work}


According to the recent study \cite{thompson2020}, which describes a review of deep learning methods for glaucoma detection, spectral domain (SD) OCT has become the most widespread diagnostic tool for analysing glaucomatous pathologies. Besides, the OCT system is routinely used in clinical practice for determining glaucoma severity since it allows emphasising the significance of the structural changes of the retina \cite{leung2012}. There are several clinical studies which claim the potential of the OCT imaging modality in the glaucoma detection paradigm. Medeiros et al. \cite{medeiros2012} enhanced the importance of the RNFL structure demonstrating that an early pathological degeneration of the retinal cells is associated with a thinning of the RNFL. In \cite{leung2012}, authors evidenced the usefulness of the RNFL thickness to determine pathological variations of the retina associated with different glaucomatous stages. Ojima et al. \cite{ojima2007} reported that the RNFL thickness has a higher potential for glaucoma diagnosis than a complete macular volume. Moreover, in \cite{abd2018}, researchers declared that RNFL features extracted from SD-OCT scans are a powerful indicator for glaucomatous damage evaluation, which is directly in line with the outcomes from our previous work \cite{garcia2020ideal}.

Inspired by the aforementioned clinical studies, many researchers have developed predictive models to discern between normal and glaucomatous patients via OCT scans analysis. Most of them focused on the optic disc due to the known potential of the circumpapillary OCT images for glaucoma detection \cite{hood2014}. In \cite{bizios2010, kim2017}, authors extracted hand-crafted features from the B-scans and combined them with transformations of A-scans and visual field parameters to discriminate between healthy and glaucoma classes. They used different machine learning classifiers such as support vector machine (SVM), random forest (RF) and \textit{k}-nearest neighbour (KNN). In \cite{ometto2019}, an automated framework to estimate the thickness of specific OCT layers was presented in order to monitor retinal abnormalities, similarly to \cite{kamal2019, hassan2019}, where the proposed methods were intended to quantify retinal layers thicknesses in search of abnormal retinal pathologies. Gao et al \cite{gao2015}, made a comparison between the thickness measurements automatically reported by the Topcon OCT system and their end-to-end framework composed of a segmentation stage followed by a feature extraction based on the RNFL thicknesses average. It is important to note that this kind of hand-driven learning methodologies usually requires a previous manual or automatic segmentation of the retinal layers to delimit the regions from which extracting the discriminative features \cite{thompson2020}. For that reason, several studies have been proposed in the literature for the sole purpose of addressing this long-standing problem \cite{niu2014, kromer2017, fang2017, duan2018, devalla2018, zang2019, pekala2019, mariottoni2020}. 

However, segmentation algorithms should entail errors which are transferred to the feature extraction stage \cite{thompson2020}. To avoid this shortcoming, deep learning models arise to provide alternative ways of quantifying structural damage since they can learn features from data automatically without reliance on previous segmentation stages or predefined features. In this context, ophthalmology has risen to a forefront in which application of deep learning (DL) algorithms can boost to a better artificial intelligence-based diagnosis in the medical fields. Specifically, glaucoma broadly meets the conditions to aid in the management of the vast amount of information coming from SD-OCT scans, according to the review outlined in \cite{thompson2020}. Following the deep learning trend, Thompson et al. \cite{thompson2020assessment} demonstrated in a recent study that their segmentation-free DL algorithm (trained with raw SD-OCT B-scans) exceeded the conventional RNFL thickness parameters to directly discern between glaucomatous and healthy eyes. Maetschke et al. \cite{maetschke2019} also conducted a comparison between hand-driven and data-driven learning strategies applied on raw OCT volumes around the ONH of the retina for glaucoma detection. The main outcomes revealed that the deep learning approach outperformed conventional SD-OCT parameters to discriminate glaucomatous from normal samples. At this point, it should be noted that there are other studies which addressed the binary classification between healthy and glaucomatous cases by applying deep learning algorithms on spectral-domain OCT volumes \cite{noury2019, wang2019, wang2020}, including our previous work \cite{garcia2020CMPB}. 

However, to the best of the authors' knowledge, the deep learning application is not very widespread on circumpapillary OCT images. A recent review of the literature found that most of the deep learning-focused studies using B-scans are conducted in a combination of fundus images to detect glaucoma via RNFL probability maps \cite{an2019, thakoor2019, shehryar2020}. To fill this gap in the literature, we previously proposed different hand-driven and data-driven learning strategies for glaucoma diagnosis in \cite{garcia2020ideal, garcia2020CMPB, garcia2020icip}, whose outcomes support the basis of this paper. 

\input{2_1_Contribution}

%% file: 2_1_Contribution.tex
\subsection{Contribution of this work}\label{subsec:2_1_Contribution}

Inspired by the high performance reported from our previous circumpapillary-based studies, we have expanded the database of B-scans around the ONH to go deeper into the glaucoma paradigm. In particular, we propose in this paper an innovative framework based on prototypical networks for glaucoma grading using raw circumpapillary images. As far as we know, this work is the first OCT-focused study intended to grade the glaucoma severity by discerning between healthy, early and advanced classes, which adds significant value to the body of knowledge. Most of the previous studies centred on B-scans are designed to discriminate glaucoma from healthy samples \cite{thompson2020assessment, asaoka2019, leung2013, iverson2014, naghizadeh2014, hammel2017}. Other state-of-the-art studies also pursued the classification of healthy and glaucomatous cases, but using the OCT volumes as an input to their models \cite{maetschke2019, noury2019, wang2019, wang2020, garcia2020CMPB}. Additional glaucoma-related studies were addressed from B-scans to accomplish different discrimination tasks such as pre-perimetric vs perimetric glaucoma \cite{na2012, na2015}, progressing vs non-progressing glaucoma \cite{na2013, wessel2013} and close angle vs open-angle glaucoma \cite{xu2019deep, fu2019deep}, among others. Furthermore, there are studies which used a different kind of input data, e.g. visual field tests \cite{li2018, kucur2018} to discern between healthy and glaucomatous patients. On a wider level, the sub-classification of early and advanced glaucoma has already been conducted in the literature, but throughout fundus image material \cite{ahn2018, zhen2018, serener2019}.


At this point, it should be noted that a very recent work \cite{raja2020} also proposes a kind of glaucoma grading set up making use of the Armed Forces Institute of Ophthalmology (AFIO) data set \cite{raja2020data}, which contains OCT scans centred on the ONH. However, it pursues the discrimination between healthy, suspects and glaucomatous samples by computing the distance of different retinal layers of interest previously segmented. In particular, Raja et al. in \cite{raja2020} propose an encoder-decoder architecture to carry out the glaucoma classification. The encoder structure was used to provide a feature map capable of discerning between healthy and glaucoma classes via softmax function; whereas the decoder component was intended to segment the interesting layers of the retina to distinguish between suspect and glaucoma cases from the samples previously predicted as glaucoma. To accomplish this part, the mean of the segmented layers was used as a feature input of an SVM classifier. Unlike AFIO database whose labels correspond to non-successive phases of the disease (healthy, suspect and glaucoma), our database includes different levels of glaucoma severity annotated according to the medical literature \cite{mills2006, susanna2009}. Contrary to Raja et al. \cite{raja2020}, we can conduct a learning framework that enables the analysis of glaucoma progression by differentiating between healthy, early and advanced glaucomatous samples. Another essential difference with respect to \cite{raja2020} is that we develop the predictive models from raw gray-scale B-scans, whereas Raja et al. \cite{raja2020} made use of pre-processed RGB scans containing manual annotations and highlighted structures. 

Furthermore, the proposed work documents additional key contributions concerning the deep learning application in the glaucoma field. For the first time, we raise a glaucoma scenario based on prototypical neural networks (PNN) \cite{snell2017}, which have demonstrated a high rate of performance in recent image analysis tasks, such as domain adaptation \cite{pan2019}, noisy evaluation \cite{gao2019}, text classification \cite{sun2019}, etc. Note that prototypical networks are usually formulated as a baseline within the few-shot paradigm \cite{fort2017, boney2017, lu2018, wang2020prototypical}, but in this paper, we exploit the prototypical concept in the \textit{k}-shot methodology to optimize the learning process for glaucoma grading.  

Tatham et al. \cite{tatham2017} argued that circumpapillary RNFL (cpRNFL) thickness was the best structure to measure glaucoma progression and the most widely used parameter in clinical practice. So, according to this, and inspired by our previous works \cite{garcia2020ideal, garcia2020icip}, we outline in this paper a novel OCT-based hybrid backbone as a feature extractor of the prototypical framework. Specifically, from \cite{garcia2020ideal}, we observed that hand-crafted features could outperform the automatic features extracted by deep learning models trained from scratch. Nevertheless, from the study carried out in \cite{garcia2020icip}, we detected that fine-tuned models also improved the model's performance compared with algorithms trained from scratch. For that reason, in this approach, we propose a novel backbone composed of pre-trained deep learning networks (with additional attention modules and residual blocks) in a combination of hand-crafted RNFL-based features, similar to the hybrid methodology that reported the best performance in \cite{garcia2020ideal}.

In summary, the main contributions of this work are:
\begin{itemize}
\item Raw circumpapillary OCT images are used for the first time to measure the glaucoma severity. 
\item Tailored prototype-based solutions are formulated in a novel framework for glaucoma grading.
\item An adapted \textit{k}-shot supervised learning, inspired by the few-shot paradigm, is conducted to exploit the specific-glaucoma knowledge.
\item A new OCT-based hybrid backbone is proposed as a feature extractor to combine automatic and hand-crafted information from B-scans. 
\end{itemize}



The rest of the paper is organized as follows. Section 3 outlines the proposed methodology composed of two stages: i) base encoder development and ii) prototype-based learning strategies. Section 4 shows the ablation experiments performed during the validation stage. Section 5 presents the quantitative and qualitative results achieved in the prediction of the test set. A wide discussion about the proposed framework is conducted in Section 6, and a summary of the main conclusions is addressed in Section 7.

%% file: 3_0_Methods.tex
\section{Methods}\label{sec:03_Methods}

\input{3_1_Encoder}

\input{3_2_Prototypes}

%% file: 3_1_Encoder.tex
\subsection{Base encoder development} \label{subsec:3_1_Encoder}

\begin{figure*}[h]
\begin{center}
\includegraphics[width=16.5cm]{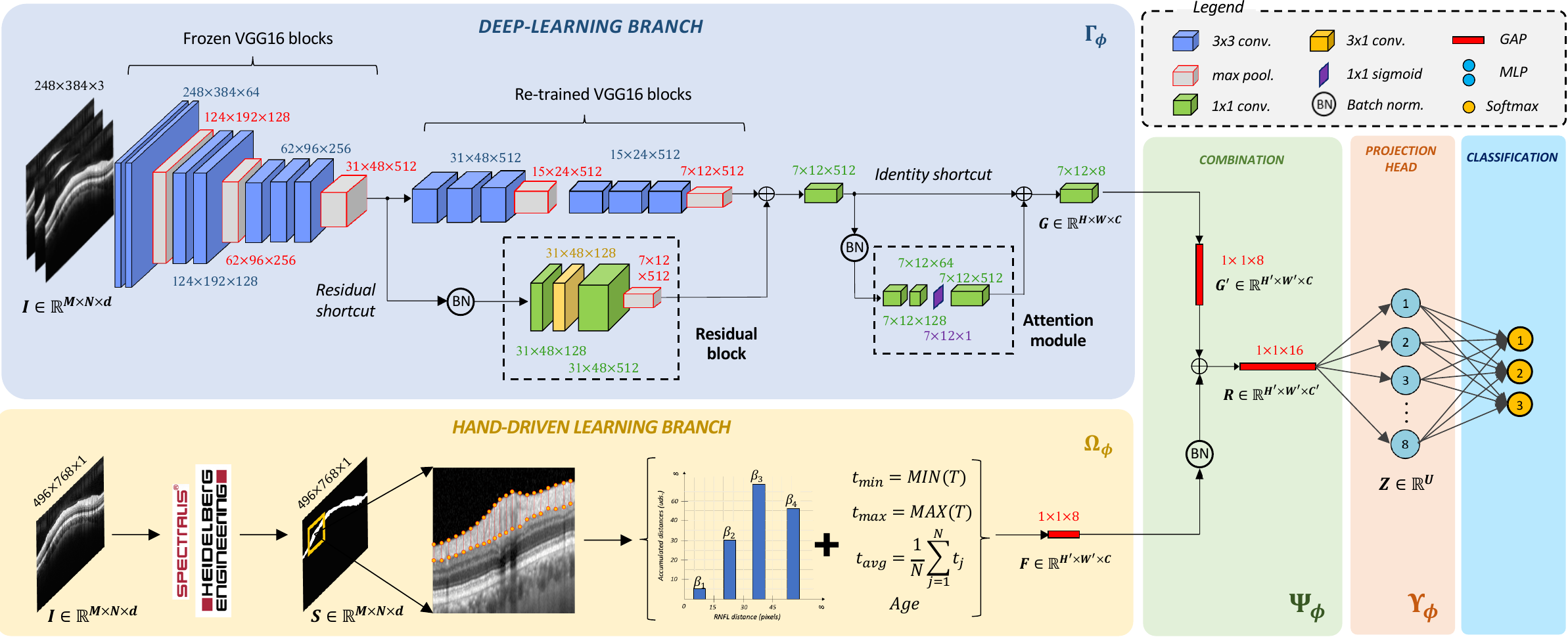}\\
\end{center}
\caption{Illustration of the end-to-end backbone proposed as a benchmark to conduct the prototype-based learning strategies. Blue, yellow and green frames correspond to the base encoder network which consists of deep learning and hand-driven learning branches followed by a combination module, respectively. A projection head (red) maps the embedded representations in a lower-dimensional space to maximize the agreement in the classification stage (cyan).}
\label{framework}
\end{figure*}

In this paper, we pay special attention to this section since the performance of the prototypical frameworks largely depends on the representation vectors encoded in the latent space by the feature extractor. The original study of the prototypical neural networks (PNN) \cite{snell2017}, as well as others derived from it \cite{fort2017, pan2019}, made use of a 4-layer CNN trained from scratch as an encoder of the feature representation. However, we observed from our previous glaucoma-based works \cite{garcia2020ideal, garcia2020icip} that deep learning models trained from scratch reported the poorest performance in comparison to fine-tuning the models or even extracting hand-crafted features. For that reason, we built on our previous experience to propose in this work a new tailored backbone able to capture the OCT-specific cues for an optimal glaucoma grading. Specifically, we inspired on the OCT hybrid methodology followed in \cite{garcia2020ideal}, but using pre-trained networks according to \cite{garcia2020CMPB, garcia2020icip}, to provide a novel base encoder $\Psi_{\phi}$ with some novelties that allow improving the previous approaches. The proposed backbone is composed of two learning branches giving rise to a multi-input feature model, as observed in Fig. \ref{framework}.

\subsubsection{Deep learning branch}
From the sweep of the pre-trained networks carried out in \cite{garcia2020icip}, we selected the VGG16 architecture as the baseline of our deep learning branch $\Gamma_{\phi}$. In particular, we applied a deep fine-tuning strategy \cite{tajbakhsh2016} to freeze the weights of the three first convolutional blocks, which were pre-trained with around 14 million of natural images corresponding to the \textit{ImageNet} data set. As a novelty, we propose a glaucoma-specific residual structure which allows propagating the information from the initial VGG layers to the last ones, via convolutional-skip connections. Thereby, the shortcut flows through the gradient of a deeper network to mitigate the problem of vanishing gradients. The proposed residual block aims to optimise the dimensionality of the filters by combining $1\times1$ convolutions (green boxes) with a customized $3\times1$ convolutional layer (yellow box). This tailored kernel size allows taking advantage of the domain-specific knowledge of the OCT images to provide local cues for the glaucoma learning process. In this way, the network is encouraged to focus on the OCT vertical axis in order to learn the glaucoma-specific information underlying the contrast differences of the retinal layers. An additional $1\times1$ convolutional layer was included to reduce the filters' dimension after concatenating the feature maps from the VGG and residual structures. At this point, we also included an attention module via skip-connections to refine the features in the spatial dimension. The proposed module works as a kind of autoencoder composed of $1\times1$ convolutions in which the filters are decreased and increased, respectively. At the bottleneck, a single convolutional filter activated by a sigmoid function (purple layer) was included to recalibrate the inputs and forcing the network to learn useful properties from the input representations. The skip-connection allows propagating larger gradients to previous layers throughout an identity shortcut. At the end of the deep learning branch (see Fig. \ref{framework}), another $1\times1$ convolutional layer was defined to provide a  volume map $G=\{g_1, g_2, ..., g_k, ..., g_C\}$, where $C=8$ is the number of feature maps $g_k$, of size $H\times W = 7\times 12$, which compose the set of the deep learning representations $\Gamma_{\phi}: I\rightarrow G$.

\subsubsection{Hand-driven learning branch}
Otherwise, inspired by the clinical study \cite{tatham2017}, we conducted an additional hand-driven learning branch $\Omega_{\phi}$ focused on hand-crafted features extracted from the retinal nerve fibre layer (RNFL). Specifically, we made use of an innovative RNFL descriptor, which was proposed in our previous work \cite{garcia2020ideal} to include RNFL thickness-based information by computing a bespoke OCT-specific histogram. Unlike \cite{mayer2010, bizios2010, kim2017, abd2018, raja2020}, among others, where descriptors were based on the RNFL mean, the proposed method allows leveraging the individual information provided by the RNFL thickness at each point of the B-scan, by considering bags of similar thicknesses values. Thus, let $I$ be a raw circumpapillary image of dimensions $M\times N \times d$, and $S$ its corresponding RNFL mask segmented automatically by the Heidelberg Spectralis OCT system, a vector of thicknesses $T=\{t_1, t_2, ..., t_j, ..., t_N\}$ was computed, where $t_j$ is the thickness value at the position $j$ of the B-scan $I$, with $j=1,2,3, ..., N$. The proposed histogram-based descriptor is able to quantify the RNFL information into $b=4$ bags depending on the thickness values. In this way, each bag $\beta_b$ collects the number of thicknesses $t_j$ whose value is ranged between $D_b$ and $D_{b+1}$, being $D=[0,15,30,45,\infty]$ a vector of relevant distances optimized on the training images. Additionally, the minimum, maximum and average of the RNFL thickness, besides the age of the patients, were also considered according to the equations formulated in Fig. \ref{framework}. Finally, the hand-driven learning branch provides a feature vector $F$ consisted of $C=8$ RNFL-specific features, such that $\Omega_{\phi}: I \rightarrow F$. 

\subsubsection{Combination module}
Once automatic and hand-crafted features were extracted from their respective branches, a simple combination module was proposed to join the embedded information in a holistic map representation $R$ composed of $C'=16$ variables per learning instance, as observed in Fig. \ref{framework}. In particular, the feature volume $G$ extracted from the deep learning branch was mapped to a vector $G'$ by a Global Average Pooling (GAP) layer, according to Equation \ref{GAP}. This operation computes a spatial squeeze from $H\times W$ to $H'\times W'$ that enables the concatenation between the features $G$ and $F$ coming from the different branches. Thereby, the proposed hybrid encoder $\Psi_{\phi}$ learns the embedding representations $R$ from the input $I$ as follows: $R=\Psi_{\phi}(I)=\Gamma_{\phi}(I)\oplus \Omega_{\phi}(I)$, where $\oplus$ denotes a concatenation operation. 

\begin{equation}
    g'_k = \frac{1}{H\times W} \sum_{h=1}^{H} \sum_{w=1}^{W} g_k(h,w)
\label{GAP}
\end{equation}

\subsubsection{Projection head module}
In this paper, we instantiate a projection head network $\Upsilon_{\phi}$ that maps the representations $R$ to an embedding vector $Z$ where the classification stage is addressed in a lower-dimensional space. The projection head $\Upsilon_{\phi}$ is comprised of a small multi-layer perceptron (MLP) with one hidden layer non-linearly activated by the ReLU function (see Fig. \ref{framework}). The use of a projection head network is widely used in very recent state-of-the-art techniques, such as contrastive learning \cite{khosla2020, chen2020, le2020}, to maximize the classification agreement. In this paper, we project the representations of the latent space via $Z = \Psi_{\phi}(R)$ to evidence that the new backbone $\Psi_{\phi}$ is better than the previous feature extractors proposed in \cite{garcia2020ideal, garcia2020CMPB, garcia2020icip}. According to the aforementioned studies \cite{khosla2020, chen2020, le2020}, the projection head network is then discarded during the prototypical learning stage to measure the distances from the representations $R=\Psi_{\phi}(I)$ to each prototype. For comparison purposes, a softmax function with three neurons was defined in the last dense layer to contrast a conventional classification approach with the proposed prototypical frameworks. 

\begin{figure*}[h]
\begin{center}
\includegraphics[width=13 cm]{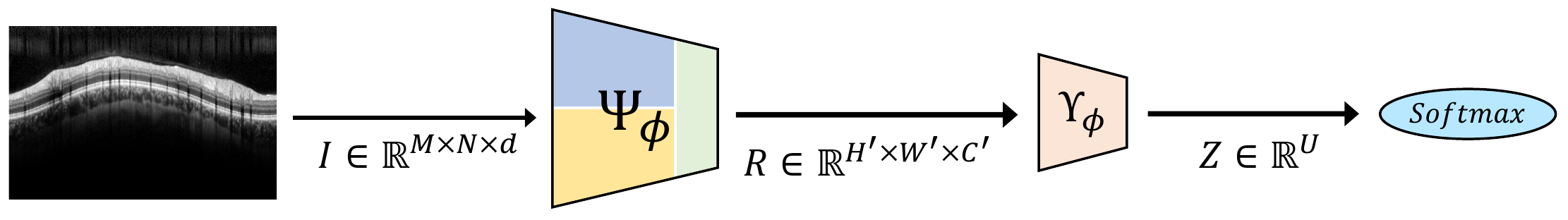}\\
\end{center}
\caption{Pipeline showing the backbone architecture composed of the base encoder $\Psi_{\phi}$, the projection head network $\Upsilon_{\phi}$ and the softmax function.}
\label{pipeline}
\end{figure*}

As a summary of the detailed base encoder backbone, we show a pipeline in Fig. \ref{pipeline} that collects the essential information. Given an input B-scan $I \in \mathbb{R}^{M\times N\times d}$, with $M\times N\times d = 248\times 384\times 3$ the dimensions of $I$, a feature embedded map $R \in \mathbb{R}^{H'\times W' \times C'}$ is provided by the encoder $\Psi_{\phi}: I \rightarrow R$. Then, the projection head module $\Upsilon_{\phi}$ maps $R$ to a metric vector $Z \in \mathbb{R}^{U}$, $\Upsilon_{\phi}: R \rightarrow Z$, where $U=8 < C'$ denotes a lower dimensional space than the latent space $R$. At the end of the convolutional network, a softmax-activated dense layer is applied to address the classification stage.

%% file: 3_2_Prototypes.tex
\subsection{Prototype-based learning strategies} \label{subsec:3_2_Prototypes}

Prototypical networks were born from the idea that there exists an embedding space in which the features from the same class cluster around a single latent group, a.k.a prototype \cite{snell2017}. In this paper, we conduct an experimental methodology to analyse the performance of two different prototype-based solutions with respect to conventional learning strategies for glaucoma grading. To this end, the traditional approach was defined according to the backbone architecture exposed in Fig. \ref{pipeline}, i.e. a base encoder network followed by a projection head module with a softmax-activated function of three neurons corresponding to healthy, early and advanced glaucomatous classes. Concerning the prototype strategies, we present below a comparison between two novel approaches -static and dynamic- that use the embedding space $R$ to determine the glaucoma severity.

\subsubsection{Static prototypes}
In this paper, we introduce the concept ``\textit{static}'' to reference the use of rigid prototypes extracted from an encoder network $\Psi_{\phi}$ whose weights $\phi$ were pre-trained in a previous stage through a conventional approach. Thereby, the frozen weights (denoted by $\Theta$) are inferred to the base encoder model to extract the embedding representations $R$ from both training $\tau$ and validation $\nu$ sets. Note that $R^\tau$ is used to obtain the prototypes and $R^\nu$ to find the nearest class prototype in the latent space. 

A similar procedure based on measuring the similarity of the latent features from different input data has been applied on different state-of-the-art techniques, such as contrastive learning \cite{khosla2020, chen2020, le2020} or content-based image retrieval (CBIR) \cite{petscharnig2017, song2017, daoud2019}. In contrastive learning tasks, an encoder network followed by a projection head is trained to differentiate between positive and negative samples, being positive samples the augmented version of a query (or samples with the same label in the case of supervised contrastive \cite{khosla2020}), and negative samples the entire remainder of the batch. Contrarily, CBIR studies train convolutional autoencoders and extract the latent space from the encoder structure to find relevant images retrieved from a set of reference that shares similar embedding features with the query image. Previous systems present some similarities with the proposed static prototype approach, since all the methods train an encoder network that is then frozen to face a second classification stage. In the case of contrastive-learning studies, the feature map extracted by the encoder architecture is used to predict the class of the query sample via MLP, k-nearest neighbours (KNN) or inferred prototypes, according to \cite{khosla2020}. In contrast, CBIR studies are intended to find the reference set samples that most closely resemble the query image by measuring the similarity of the embedded representations extracted by the encoder structure. The proposed approach differs from the previous ones in a critical point: the encoder network is trained during the first stage for the same objective to be achieved in the second one, i.e. glaucoma grading. Oppositely, contrastive-learning and CBIR-based studies use backbones that were pre-trained for a different task during the first stage. 

\begin{figure*}[h]
\begin{center}
\includegraphics[width=13cm]{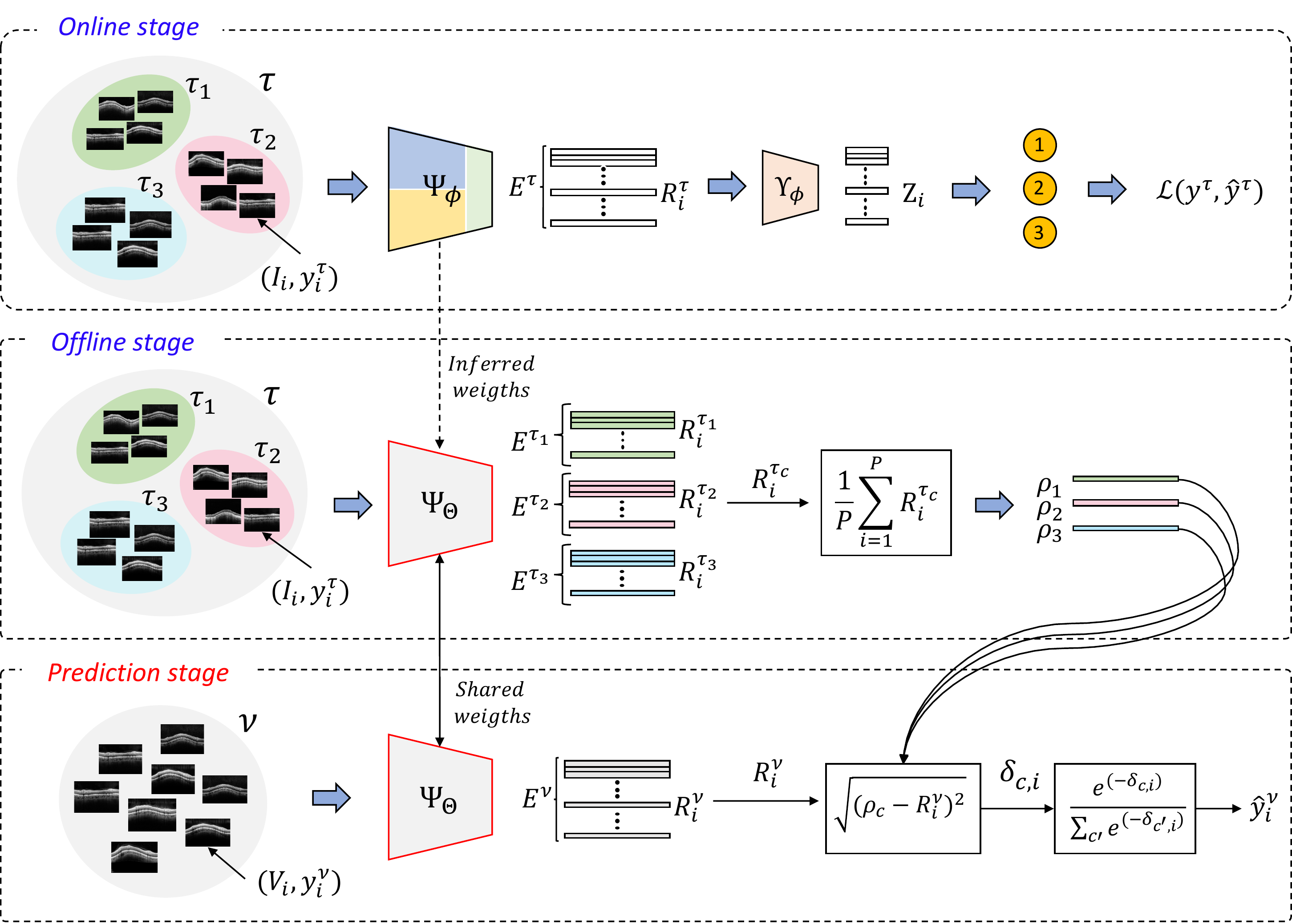}\\
\end{center}
\caption{Proposed static prototype-based learning strategy. A conventional approach is conducted in the online stage to optimize the encoder. In the offline stage, the weights of the pre-trained backbone are inferred to extract a prototype $\rho_c$ per class. In the prediction phase, the class of a validation B-scan $V_i$ is determined by measuring the latent distance between each prototype $\rho_c$ and the embedding representations $R_i^{\nu} = \Psi_{\phi}(V_i)$.}
\label{flowchart_static_method}
\end{figure*}

Below, we detail the training of the proposed approach based on static prototypes, which is composed of two (online and offline) stages, according to Fig. \ref{flowchart_static_method}. 

In the online stage, a conventional classification pipeline was conducted to optimize the weights of the proposed OCT-hybrid backbone by minimizing the categorical cross-entropy loss function $\mathcal{L}(y^\tau,\hat{y}^\tau)$ in each training epoch $e=1, 2, 3, ..., \epsilon$, as detailed in Algorithm \ref{static_prototype}.
In the offline stage, the weights of the pre-trained encoder $\Psi_{\phi}$ were frozen ($\phi \xrightarrow{} \Theta$) and the projection head $\Upsilon_{\phi}$ and softmax modules were discarded to avoid the non-linearity of the top model. The embedding representations $E^\tau=\{R_1, R_2, ..., R_i, ..., R_{P^\tau}\}$, with $P^\tau$ the number of samples of the minority class in the training set $\tau$, were used to infer the rigid prototypes. In particular, each $\rho_c$ was calculated as the mean of the latent representations $R_i^{\tau_c}=\Psi_{\phi}(I_i)$, where $I_i\in\tau_c\subset\tau$ denotes the \textit{i}-sample of the training set $\tau$ associated with the class $c$ (see Equation \ref{mean_prototype}). 

\begin{equation}
    \rho_c = \frac{1}{P^{\tau_c}}{\sum_{i=1}^{P^{\tau_c}}{R_i^{\tau_c}}}
\label{mean_prototype}
\end{equation}

During the prediction phase, a matrix of distances $\delta_{c,i}$ was achieved by measuring the Euclidean distance (Equation \ref{euclidean}) between each prototype $\rho_c$ and the embedding representation $R_i^{\nu}=\Psi_{\phi}(V_i)$, where $V_i\in\nu$ corresponds to the \textit{i}-scan of the validation subset $\nu$. From here, a probability of belonging to each class was calculated (Equation \ref{probability}) to determine the predicted class $\hat{y}_i^\nu$, as detailed in Algorithm \ref{static_prototype}. 

\begin{equation}
    \delta_{c,i} = \sqrt{(\rho_c-R_i^{\nu})^2}
\label{euclidean}
\end{equation}

\begin{equation}
    p_{i,c} = \frac{\exp{(-\delta_{c,i})}}{\sum_{c'}{\exp{(-\delta_{c',i})}}}
\label{probability}
\end{equation}

\begin{algorithm}[h!]
\caption{Static prototype-based strategy.}
\label{static_prototype}
\footnotesize
\small
\BlankLine
\KwData{Training $\tau=\{(I_1,y_1^\tau), ..., (I_{P^\tau},y_{P^\tau}^\tau)\}$ and validation $\nu=\{(V_1,y_1^\nu), ..., (V_{P^\nu},y_{P^\nu}^\nu)\}$ sets.}
\textbf{Results:}\\
\textit{Online stage} $\leftarrow$ Trained base encoder $\Psi_\phi$ \;
\textit{Offline stage} $\leftarrow$ Inferred prototypes $\rho_c$ \;
\textit{Prediction stage} $\leftarrow$ Predicted labels $\hat{y}_i^\nu$.
\BlankLine
\textbf{Algorithm:} \\
\textit{Online stage: }\\
$\phi \leftarrow $ random\;
\For{$ e \leftarrow 1$ \KwTo $\mathbf \epsilon$}{
    \For{$ i \leftarrow 1$ \KwTo $\mathbf P^{\tau}$}{
        $R_i^{\tau} \leftarrow \Psi_{\phi}(I_i)$ \;
        $Z_i \leftarrow \Upsilon_{\phi}(R_i^{\tau})$ \;
        $\hat{y}_i^\tau \leftarrow$ softmax$(Z_i)$ \;
    }
    $\mathcal{L}(y^\tau,\hat{y}^\tau) \leftarrow -\sum_{i}{y_i^\tau~log(\hat{y}_i^\tau)}$ \;
    Update $\phi$ using $\nabla_\phi \mathcal{L}$ \;
}
\BlankLine
\textit{Offline stage:} \\
\For{$ c \leftarrow 1$ \KwTo $\mathbf 3$}{
    $\rho_c \leftarrow \frac{1}{P^{\tau}}{\sum_{i=1}^{P^{\tau}}{\Psi_{\Theta}(I_i)}} $ \;
}
\BlankLine
\textit{Prediction phase: }\\
\For{$ i \leftarrow 1$ \KwTo $\mathbf P^{\nu}$}{
    $R_i^{\nu} \leftarrow \Psi_{\Theta}(V_i)$ \;
    \For{$ c \leftarrow 1$ \KwTo $\mathbf 3$}{
        $\delta_{i,c} \leftarrow \sqrt{(\rho_c-R_i^{\nu})^2} $ \;
        $p_{i,c} \leftarrow \frac{\exp{(-\delta_{i,c})}}{\sum_{c'}{\exp{(-\delta_{i,c'})}}} $ \;
    }
    $\hat{y}^\nu_i \leftarrow argmax(p_{i,c})$
}
\end{algorithm}

\subsubsection{Dynamic prototypes} \label{3_2_2_Dynamic_prototypes}
\begin{figure*}[h]
\begin{center}
\includegraphics[width=16cm]{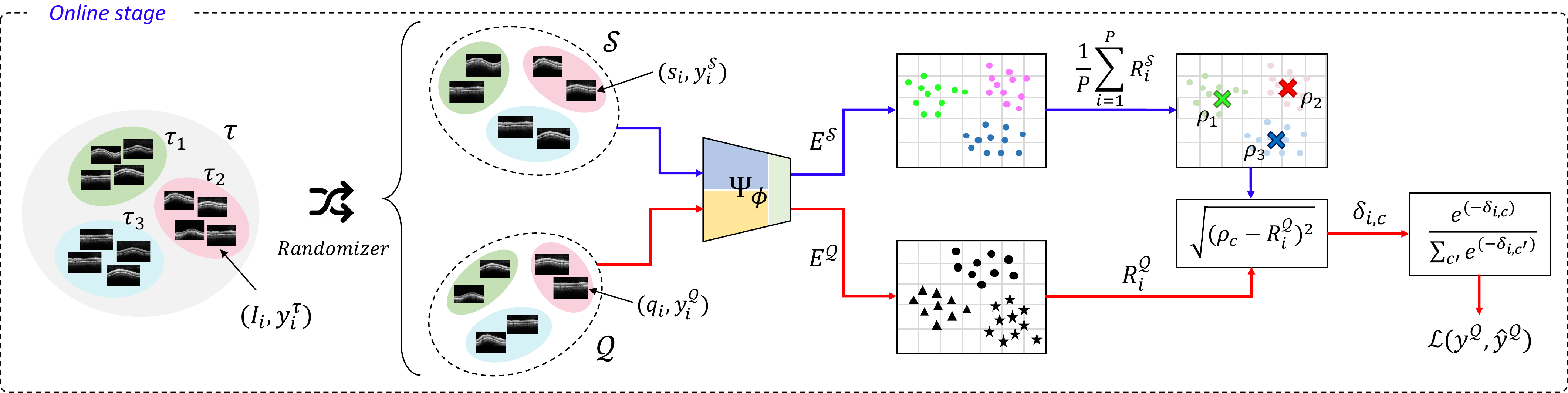}\\
\end{center}
\caption{Flowchart of the online stage corresponding to the dynamic prototype-based learning strategy. A support set $\mathcal{S}$ and a query set $\mathcal{Q}$ are randomly selected from the training set $\tau$ to develop a supervised PNN. The hybrid OCT-based backbone $\Psi_\phi$ is used to extract the embedding representations from both $\mathcal{S}$, to determine the prototypes $\rho_c$, and $\mathcal{Q}$, to map the latent representations from the query samples. At the end of the training process, a softmax function is applied to predict the query label based on the latent distances $\delta_{i,c}$.}
\label{flowchart_dynamic_method}
\end{figure*}

Inspired by \cite{snell2017}, where prototypical neural networks (PNNs) were proposed for few-shot learning, we present in this paper a PNN-based framework for grading glaucoma by exploiting the \textit{k}-shot methodology. The main difference with respect to the previous static approach lies in the online stage since dynamic prototypes are trained in an end-to-end manner, such that prototypes are updated after each epoch $e$. In this way, the base encoder network can be optimized according to latent distances, instead of a conventional classification top model, as observed in Fig. \ref{flowchart_dynamic_method}.

From here on, we will employ the terminology ($\mathcal{N}$-way, $\mathcal{K}$-shot) used in the literature for few-shot learning, where $\mathcal{K}$ is the number of labelled samples and $\mathcal{N}$ the number of classes in the training set \cite{snell2017}. In the state-of-the-art studies based on prototypical networks \cite{snell2017, simon2018, simon2020}, a labelled support set $\mathcal{S}=\{(s_1,y_1), ..., (s_i,y_i), ..., (s_\mathcal{K}, y_\mathcal{K})\}$ and an unlabelled query set $\mathcal{Q}=\{q_1, q_2, ..., q_i, ..., q_\mathcal{U}\}$, are considered to train the PNN-based models in a few-shot scenario, being $\mathcal{U}$ the number of unlabelled samples selected from the training set $\tau$. Specifically, $\mathcal{S}$ is used to extract each of $\mathcal{N}$ prototypes $\rho_c$, whereas $\mathcal{Q}$ is employed to find the nearest class prototype for an embedded query point $R_i^\mathcal{Q}=\Psi_\phi(q_i)$. A negative log-probability loss is updated according to the distance metrics from $\rho_c$ and $R_i^\mathcal{Q}$.

The proposed dynamic prototype-based learning strategy differs from the state of the art in multiple ways. Unlike aforementioned studies \cite{snell2017, simon2018, simon2020}, where authors selected a specific number of $\mathcal{U}=5,10$ or $15$ unlabelled samples, depending on the database addressed, in this study we made use of $\mathcal{U}=P-\mathcal{K}$ labelled query samples to get the most out of the training set $\tau$. Note that we propose, for the first time, the use of a labelled query set $\mathcal{Q}$, giving rise to a novel Supervised Prototypical Neural Network for Glaucoma Grading (SPNN-GG). This results in another difference concerning the literature, since a supervised approach allows including the ground truth label $y^\mathcal{Q}$ in the loss function, so learning proceeds by minimizing the categorical cross-entropy, instead of a log-probability function. Also, as a novelty, we move away from the few-shot setting by proposing an optimal $k-$shot scenario in which $\mathcal{K} \in [1, P-\mathcal{U}]$ is optimized as an additional hyper-parameter, unlike in the previous studies where $\mathcal{K} \in [1,5]$.

The online training of the proposed dynamic prototype-based approach is conducted in a supervised end-to-end way, according to Fig. \ref{flowchart_dynamic_method}. It is important to remark that the main difference between static and dynamic prototypes lies here since the static approach infers rigid prototypes from a backbone trained by a projection head module, whereas the dynamic strategy updates the prototypes during the training phase by optimizing a backbone based on latent distances. Also note that offline and prediction phases are addressed as in the static approach, i.e. using the entire training set $\tau$ to extract the prototypes $\rho_c$, as detailed in Algorithm \ref{dynamic_prototype}.

From the training set $\tau$ containing $P_c^\tau$ samples for each class $c$, a support set $\mathcal{S}=\{(s_1,y_1^\mathcal{S}), ..., (s_\mathcal{K}, y_\mathcal{K}^\mathcal{S})\}$ and a query set $\mathcal{Q}=\{(q_1,y_1^\mathcal{Q}), ..., (q_\mathcal{U}, y_\mathcal{U}^\mathcal{Q})\}$, with $\mathcal{U}=P-\mathcal{K}$, are obtained after a randomization process at every epoch $e$. The hybrid OCT-based backbone $\Psi_\phi$ was used as the encoder network to extract the embedding representations from the support $s_i$ and query $q_i$ samples. Specifically, latent support set-coming features $E^\mathcal{S}=\Psi_\phi(\mathcal{S})=\{R_1^\mathcal{S}, ..., R_i^\mathcal{S}, ... R_\mathcal{K}^\mathcal{S}\}$ are used to extract each class-prototype $\rho_c$ as the mean of the embedded representations $R_i^\mathcal{S}=\Psi_\phi(s_i)$. In contrast, query representations $E^\mathcal{Q}=\Psi_\phi(\mathcal{Q})=\{R_1^\mathcal{Q}, ..., R_i^\mathcal{Q}, ... R_\mathcal{U}^\mathcal{Q}\}$ are mapped in the latent space to find the closest prototype $\rho_c$, in terms of Euclidean distance. Then, a softmax function is applied to determine the probability of belonging to each class $p(\hat{y}_i^\mathcal{Q}=c|R_i^\mathcal{Q})$. During the backward-propagation step, the embedding representations are refined by updating the weights of the base encoder network at every epoch, according to the categorical cross-entropy loss function, denoted by $\mathcal{L}(y^\mathcal{Q},\hat{y}^\mathcal{Q})$. In this way, prototypes are optimized under the hypothesis that each class can be described by just one subspace. Therefore, learning progresses by minimizing the latent distances between $R_i^{\mathcal{Q}_c \subset \tau_c}$ and $\rho_c$, unlike in the static approach where the encoder was updated according to the embedding representations $Z=\Upsilon(\Psi(I_i))$ extracted from the projection network.

\begin{algorithm}[h!]
\caption{Dynamic prototype-based learning strategy.}
\label{dynamic_prototype}
\footnotesize
\small
\BlankLine
\KwData{Training $\tau=\{(I_1,y_1^\tau), ..., (I_{P^\tau},y_{P^\tau}^\tau)\}$ and validation $\nu=\{(V_1,y_1^\nu), ..., (V_{P^\nu},y_{P^\nu}^\nu)\}$ sets.}
\textbf{Results:}\\
\textit{Online stage} $\leftarrow$ Trained base encoder $\Psi_\phi$ \;
\textit{Offline stage} $\leftarrow$ Refined prototypes $\rho_c$ \;
\textit{Prediction stage} $\leftarrow$ Predicted labels $\hat{y}_i^\nu$.
\BlankLine
\textit{Online stage: }\\
$\phi \leftarrow $ random\;
\For{$ e \leftarrow 1$ \KwTo $\mathbf \epsilon$}{
    \For{$ i \leftarrow 1$ \KwTo $\mathcal{K}$}{
        $(s_i, y_i^\mathcal{S}) \leftarrow (I_{random(i)}, y_{random(i)}^\tau) $ \;
        $R_i^\mathcal{S} \leftarrow \Psi_\phi(s_i) $ \;
    }
    \For{$ i \leftarrow 1$ \KwTo $\mathcal{U}$}{
        $(q_i, y_i^\mathcal{Q}) \leftarrow (I_{random(i)}, y_{random(i)}^\tau) \notin \mathcal{S}$ \;
        $R_i^\mathcal{Q} \leftarrow \Psi_\phi(q_i)$ \;
        \For{$ c \leftarrow 1$ \KwTo $\mathcal{N}$}{
            $\rho_c \leftarrow \frac{1}{\mathcal{K}}{\sum_{i=1}^{\mathcal{K}}{R_i^\mathcal{S}}} $ \;
            $\delta_{i,c} \leftarrow \sqrt{(\rho_c-R_i^\mathcal{Q})^2} $\;
            $p_{i,c} \leftarrow \frac{\exp{(-\delta_{i,c})}}{\sum_{c'}{\exp{(-\delta_{i,c'})}}} $ \;
        }
        $\hat{y}_i^\mathcal{Q} \leftarrow argmax(p_{i,c})$
    }
    $\mathcal{L}(y^\mathcal{Q},\hat{y}^\mathcal{Q}) \leftarrow -\sum_{u}{y_i^\mathcal{Q}~log(\hat{y}_i^\mathcal{Q})}$ \;
    Update $\phi$ using $\nabla_\phi \mathcal{L}$ \;
}
\BlankLine
\textit{Offline stage:} \\
\For{$ c \leftarrow 1$ \KwTo $\mathbf \mathcal{N}$}{
    $\rho_c \leftarrow \frac{1}{P^{\tau}}{\sum_{i=1}^{P^{\tau}}{\Psi_{\Theta}(I_i)}} $ \;
}
\BlankLine
\textit{Prediction phase: }\\
\For{$ i \leftarrow 1$ \KwTo $\mathbf P^{\nu}$}{
    $R_i^{\nu} \leftarrow \Psi_{\Theta}(V_i) $ \;
    \For{$ c \leftarrow 1$ \KwTo $\mathbf 3$}{
        $\delta_{i,c} \leftarrow \sqrt{(\rho_c-R_i^{\nu})^2} $ \;
        $p_{i,c} \leftarrow \frac{\exp{(-\delta_{i,c})}}{\sum_{c'}{\exp{(-\delta_{i,c'})}}} $ \;
    }
    $\hat{y}_i^\nu \leftarrow argmax(p_{i,c})$ \;
}
\end{algorithm}

%% file: 4_0_Ablation_experiments.tex
\section{Ablation experiments}\label{sec:04_Ablation_experiments}

\input{4_1_Datasets}

\input{4_2_Backbone_selection}

\input{4_3_Prototype_strategies}

%% file: 4_1_Datasets.tex
\subsection{Data sets} \label{subsec:4_1_Datasets}

Two private databases coming from different sources were used to develop and evaluate the predictive models for glaucoma grading. Both data sets (\textit{Database 1} and \textit{Database 2}) contain high-resolution SD-OCT scans, which were acquired from healthy, early and advanced-glaucomatous patients with an axial resolution of 3-4 $\mu m$ using the Heidelberg Spectralis OCT system. This equipment provides circumpapillary B-scans centred on the optical nerve head (ONH) of the retina throughout a super-luminescence diode with an infrared beam of an average wavelength of 870 nm and a bandwidth of 25 nm. In particular, the samples were extracted with a resolution of $496\times 768$ pixels.

Note that subjects with open-angle glaucoma (POAG) were included in the study, whereas patients suffering other ocular disorders, such as closed-angle glaucoma or pseudoexfoliation syndrome were discarded from both databases. Also, patients with media opacity were excluded if the opacity disturbs the B-scan OCT imaging critically. Thus, B-scans with a poor-quality OCT image were discarded from the study. Two different senior ophthalmologists (with more than 25 years of professional experience in clinical ophthalmology) carried out the annotation of the databases. Specifically, each expert manually labelled just one data set, following the European guideline for Glaucoma diagnosis. The examination included several tests such as Goldman applanation tonometry, gonioscopy, slit lamp examination, standard automated perimetry and thickness measurement of specific retinal layers of interest. According to the clinical literature \cite{mills2006, susanna2009}, the mean deviation (MD) score plays an essential role in the glaucoma grading scale, such that the severity of the glaucoma depends on the range in which the MD value is found. Conforming to \cite{mills2006}, $MD>=-6dB$ is Early; $-6dB>MD\geq-12dB$ is Moderate; $-12dB>MD\geq-20dB$ is Advanced and $MD<-20dB$ is Severe. Since our objective is contributing to the glaucoma grading just from OCT images, without performing additional time-consuming tests, we simplify the glaucoma staging scale by labelling as Advanced those glaucomatous samples with an $MD<-6dB$. Note that the proposed system is not intended to serve as a definitive glaucoma diagnosis, but as a diagnositc tool which allows guiding the expert's decision through approximate but reliable OCT-based results. Based on the above analyses, all B-scans were classified as healthy, early or advanced. In Tables \ref{dataset_distribution} and \ref{age_gender}, we show more information about the specifications of each data set. 

\begin{table}[h]
\caption{Number of patients (pat.) and samples (samp.) in each database grouped by categories, according to the experts' annotation.}
\label{dataset_distribution}
\renewcommand{\arraystretch}{1} 
\setlength\tabcolsep{5 pt} 
\centering
\resizebox{8cm}{!}{
\begin{tabular}{cccc|c}
\hline
                    & \begin{tabular}[c]{@{}c@{}}\textbf{Healthy}\\ \textit{(pat./samp.)}\end{tabular} & \begin{tabular}[c]{@{}c@{}}\textbf{Early}\\ \textit{(pat./samp.)}\end{tabular} & \begin{tabular}[c]{@{}c@{}}\textbf{Advanced}\\ \textit{(pat./samp.)}\end{tabular} & \begin{tabular}[c]{@{}c@{}}\textbf{TOTAL}\\ \textit{(pat./samp.)}\end{tabular} \\ \hline
\textbf{Database 1} & 32 / 41       & 28 / 35     & 25 / 31      & 85 / 107     \\
\textbf{Database 2} & 26 / 49       & 24 / 37     & 21 / 26      & 71 / 112     \\ \hline
\textbf{TOTAL}      & 58 / 90       & 52 / 72     & 46 / 57      & 156 / 219    \\ \hline
\end{tabular}}
\end{table}

\begin{table}[h]
\caption{Additional information about the age and gender of the patients who compose each database.}
\label{age_gender}
\renewcommand{\arraystretch}{1} 
\setlength\tabcolsep{5 pt} 
\centering
\resizebox{8cm}{!}{
\begin{tabular}{ccccc}
\hline
                    & \multicolumn{2}{c}{\textbf{Age}} & \multicolumn{2}{c}{\textbf{Gender}} \\ \cline{2-5} 
                    & \textit{Range}   & \textit{$\mu\pm\sigma$}  & \textit{Male}   & \textit{Female}   \\ \hline
\textbf{Database 1} & {[}19-88{]}      & 60,45$\pm$16,54   & 46 (54,12\%)    & 39 (45,88\%)      \\
\textbf{Database 2} & {[}30-90{]}      & 64,80$\pm$13,93   & 26 (36,62\%)    & 45 (63,38\%)      \\ \hline
\textbf{TOTAL}      & {[}19-90{]}      & 62,44$\pm$15,51   & 72 (46,15\%)    & 84 (53,85\%)      \\ \hline
\end{tabular}}
\end{table}

It is important to note that, as claimed in \cite{raja2020}, there are no public glaucoma-labelled OCT databases that enable an objective comparison with our work. To the best of the authors' knowledge, Armed Forces Institute of Ophthalmology (AFIO) data set \cite{raja2020data} is the only publicly available repository of ONH SD-OCT scans of healthy and glaucomatous subjects. However, the differences between those B-scans and ours make the direct application of our algorithms impossible for multiple reasons: i) B-scans from AFIO data set present manual annotations and highlighted structures of interest. ii) OCT images have been pre-processed showing an RGB colour mode in which cup-to-disk regions appears remarked. iii) AFIO database was acquired using Topcon 3D OCT-1000 machines. iv) The experts' annotations include healthy, glaucoma and suspect labels. Differently, the databases used here contain the gray-scale OCT samples extracted in raw by the Heidelberg Spectralis OCT equipment, without any manual annotation or pre-processing. In addition, our database is explicitly labelled into healthy, early and advanced glaucoma classes, according to the visual field-based criteria of the medical literature \cite{mills2006, susanna2009}.  

Furthermore, another OCT database (OCTID) (with raw ONH B-scans similar to ours) is publicly available in \cite{gholami2020}. However, OCTID data set includes Normal (NO), Macular Hole (MH), Age-related Macular Degeneration (AMD), Central Serous Retinopathy (CSR) and Diabetic Retinopathy (DR) image classes, but it does not contemplate the glaucoma class. 

\BlankLine
\textbf{Data partitioning}.
In this paper, we fuse \textit{Database 1} and \textit{Database 2} in order to increase the number of samples from which to develop our machine learning algorithms and to evidence the reliability of the predictive models using two data sets coming from different sources. Making use of the entire fused database, we conducted a patient-level data partitioning procedure to separate training and testing sets. Specifically, $\frac{1}{6}$ of the data was used to test the models, whereas the remainder of the database was employed to train the algorithms. From the training set, we randomly split the data again into training and validation subsets, according to Table \ref{partition}, to optimise the models' hyper-parameters and monitor the over-fitting. 

\begin{table}[h]
\caption{Partition of the circumpapillary B-scans to develop and evaluate the predictive models.}
\label{partition}
\renewcommand{\arraystretch}{1.1} 
\setlength\tabcolsep{6 pt} 
\centering
\resizebox{6cm}{!}{
\begin{tabular}{cccc}
\hline
\multicolumn{1}{l}{} & \textbf{Healthy} & \textbf{Early} & \textbf{Advanced} \\ \hline
\textbf{Training}    & 60               & 48             & 41                \\
\textbf{Validation}  & 15               & 12             & 10                \\
\textbf{Test}        & 15               & 12             & 6                 \\ \hline
\end{tabular}}
\end{table}

%% file: 4_2_Backbone_selection.tex
\subsection{Backbone selection} \label{subsec:4_2_Backbone_selection}

Unlike most of the state-of-the-art studies which used as a feature extractor either well-known architectures such as ResNet or VGG \cite{simon2020, khosla2020} or simpler CNNs trained from scratch \cite{snell2017, pan2019}, we pretend to exploit the feature extraction stage to get the most out from the circumpapillary OCT scans. For this reason, we propose a novel OCT-hybrid backbone inspired by ophthalmic clinical studies \cite{tatham2017} and our previous glaucoma detection-based experience applying hand-driven \cite{garcia2020ideal} and deep learning algorithms \cite{garcia2020icip, garcia2020CMPB}. To address an objective comparison with other state-of-the-art studies, we contrast in Tables \ref{validation_perclass} and \ref{validation_averages} the validation results achieved by different architectures trained in a multi-class scenario. Particularly, we compared four approaches, as detailed below.

\begin{enumerate}
    \item \textit{RNFL features}. A simple MLP was trained using the output of the hand-driven learning branch $\Omega_\phi(I)$ as an input data, similarly to \cite{garcia2020ideal}.
    \item \textit{Fine-tuned VGG16}. This very popular architecture was used (freezing the three first convolutional blocks) as a feature extractor followed by the same MLP classifier as before, according to \cite{garcia2020icip}. 
    \item \textit{Fine-tuned RAGNet}. An expanded version of the previous approach was conducted by including residual and attention modules in the feature extraction architecture. This approach, corresponding to the deep learning branch $\Gamma_\phi(I)$, was introduced in our previous work \cite{garcia2020CMPB}.
    \item \textit{OCT-hybrid network}. This approach corresponds to the end-to-end backbone $\Psi_\phi(I)$ exposed in Fig. \ref{framework}, which proposes a combination of hand-crafted and automatic features before the top model. 
\end{enumerate}

The comparison was handled by means of different figures of merit, such as sensitivity (SN), specificity (SP), positive predictive value (PPV), negative predictive value (NPV), F-score (FS) and accuracy (ACC). Notably, the backbone reporting the best performance during the validation stage was selected as the base encoder network to address the next prototypical-based learning strategy. 

\BlankLine
\textbf{Training details.} All the contrasting approaches were implemented using Tensorflow 2.3.1 on Python 3.6. Experiments were conducted on a machine with Intel(R) Core(TM) i7-9700 CPU @3.00GHz processor and 16GB RAM. A single NVIDIA GeForce RTX 2080 having cuDNN 7.5 and a CUDA Toolkit 10.1 was used to develop the deep learning algorithms. All models were trained during 200 epochs using a learning rate of 0.0005 with a batch size of 16. Stochastic gradient descent (SGD) optimizer was applied trying to minimize the categorical cross-entropy (CCE) loss function at every epoch. The rest of the architecture hyper-parameters and input dimensions are shown in Fig. \ref{framework}.

%% file: 4_3_Prototype_strategies.tex
\subsection{Prototype-based learning strategies} \label{subsec:4_3_Prototype_strategies}

In this section, we report the validation performance of the static and dynamic prototype-based methods in comparison to the conventional multi-class approach. It is important to note that the comparison was conducted using the proposed OCT-hybrid backbone as a feature extractor for all the scenarios. Following the organization of the previous section, Tables \ref{val_proto_perclass} and \ref{val_proto_averages} show the comparison between the three involved learning strategies during the validation phase. 

\begin{figure}[t]
\begin{center}
\includegraphics[width=7cm]{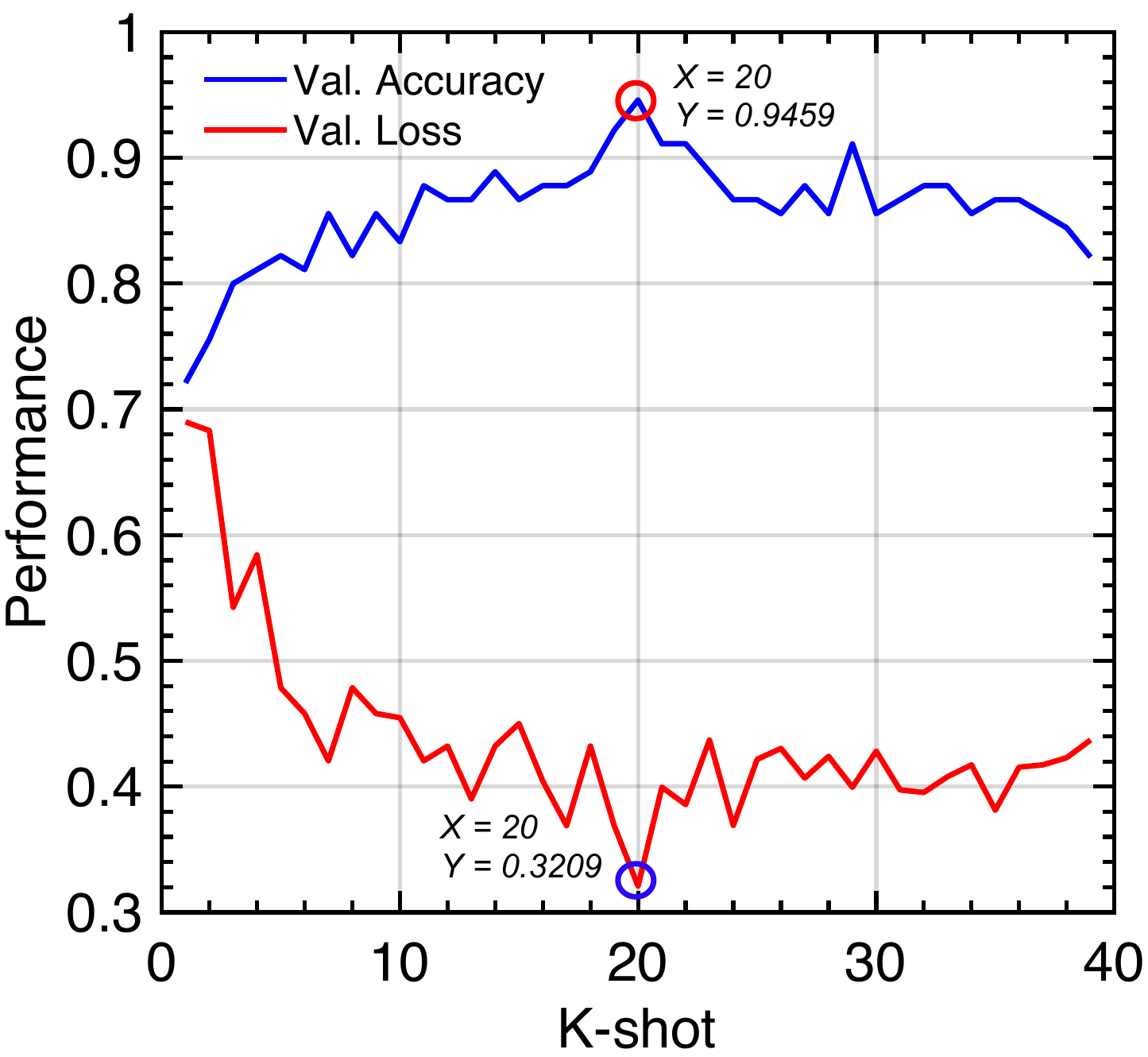}\\
\end{center}
\caption{Models' performance using $\mathcal{K}$ samples from each class $c$ to define the prototypes $\rho_c$ and $\mathcal{U}=P-\mathcal{K}$ query samples to measure the embedding distances to minimize the loss function.}
\label{val_shot}
\end{figure}

\textbf{Training details.} The same hardware and software systems as before were used to accomplish this section. However, some differences in the dynamic prototypical approach are worth noting. Dimensions of the input image must be downsized to $124\times192\times3$ to face the GPU memory constraints. Additionally, a decreased learning rate of 0.001 allowed the convergence of the model in 50 training epochs. A batch size of 16 samples was defined to minimize the CEE loss function using the SGD optimizer. Regarding the specific parameters of the dynamic prototype-based strategy, the number of $\mathcal{K}$ shots and the number of $\mathcal{U}$ query samples were determined after an optimization process, according to Fig. \ref{val_shot}. Specifically, $\mathcal{K}=20$ support samples were selected to extract the prototypes $\rho_c$ as the mean of the embedding representations $E^\mathcal{S}$, whereas $\mathcal{U}=21$ query samples were used to measure the Euclidean distance between the latent features $E^\mathcal{Q}$ and each $\rho_c$. Note that $P=41$ denotes the number of training samples of the minority class (see Table \ref{dataset_distribution}). Also, other statistics and distance metrics were considered during the optimization of the models, as observed in Table \ref{distance_table}. The rest of the hyper-parameters related to the dynamic prototypes is detailed in Section \ref{3_2_2_Dynamic_prototypes}. 

\input{4_4_val_tables}

%% file: 4_4_val_tables.tex
\begin{table*}[h]
\caption{Backbone selection: Validation results per class achieved from different architectures in a multi-class scenario.}
\label{validation_perclass}
\centering
\renewcommand{\arraystretch}{1.3}
\resizebox{16cm}{!}{
\begin{tabular}{ccccc|cccc|cccc}
\hline
                                  & \multicolumn{4}{c|}{\textbf{HEALTHY}}                                                                                                                                                                                                                                                  & \multicolumn{4}{c|}{\textbf{EARLY GLAUCOMA}}                                                                                                                                                                                                                                                    & \multicolumn{4}{c}{\textbf{ADVANCED GLAUCOMA}}                                                                                                                                                                                                                                                  \\ \hline
                                  & \textit{\begin{tabular}[c]{@{}c@{}}RNFL \\ features \cite{garcia2020ideal}\end{tabular}} & \textit{\begin{tabular}[c]{@{}c@{}}Fine-tuned\\ VGG16 \cite{garcia2020icip}\end{tabular}} & \textit{\begin{tabular}[c]{@{}c@{}}Fine-tuned\\ RAGNet \cite{garcia2020CMPB}\end{tabular}} & \textit{\begin{tabular}[c]{@{}c@{}}OCT-hybrid\\ network\end{tabular}} & \textit{\begin{tabular}[c]{@{}c@{}}RNFL \\ features \cite{garcia2020ideal}\end{tabular}} & \textit{\begin{tabular}[c]{@{}c@{}}Fine-tuned\\ VGG16 \cite{garcia2020icip}\end{tabular}} & \textit{\begin{tabular}[c]{@{}c@{}}Fine-tuned\\ RAGNet \cite{garcia2020CMPB}\end{tabular}} & \textit{\begin{tabular}[c]{@{}c@{}}OCT-hybrid\\ network\end{tabular}} & \textit{\begin{tabular}[c]{@{}c@{}}RNFL \\ features \cite{garcia2020ideal}\end{tabular}} & \textit{\begin{tabular}[c]{@{}c@{}}Fine-tuned\\ VGG16 \cite{garcia2020icip}\end{tabular}} & \textit{\begin{tabular}[c]{@{}c@{}}Fine-tuned\\ RAGNet \cite{garcia2020CMPB}\end{tabular}} & \textit{\begin{tabular}[c]{@{}c@{}}OCT-hybrid\\ network\end{tabular}} \\ \hline
\multicolumn{1}{c|}{\textbf{SN}}  & \textbf{1}                                                        & \textbf{1}                                                          & 0.9333                                                               & \textbf{1}                                                            & 0.6667                                                            & 0.5833                                                              & \textbf{0.7500}                                                        & \textbf{0.7500}                                                         & 0.8000                                                               & \textbf{0.9000}                                                        & \textbf{0.9000}                                                         & \textbf{0.9000}                                                          \\
\multicolumn{1}{c|}{\textbf{SP}}  & 0.8636                                                            & 0.8636                                                              & \textbf{0.9545}                                                      & 0.9091                                                                & 0.9200                                                              & \textbf{0.9600}                                                                & 0.9200                                                                 & \textbf{0.9600}                                                         & \textbf{0.9630}                                                            & 0.9259                                                              & 0.9259                                                               & \textbf{0.9630}                                                        \\
\multicolumn{1}{c|}{\textbf{PPV}} & 0.8333                                                            & 0.8333                                                              & \textbf{0.9333}                                                      & 0.8824                                                                & 0.8000                                                               & 0.8750                                                               & 0.8182                                                               & \textbf{0.9000}                                                          & 0.8889                                                            & 0.8182                                                              & 0.8182                                                               & \textbf{0.9000}                                                          \\
\multicolumn{1}{c|}{\textbf{NPV}} & \textbf{1}                                                        & \textbf{1}                                                          & 0.9545                                                               & \textbf{1}                                                            & 0.8519                                                            & 0.8276                                                              & 0.8846                                                               & \textbf{0.8889}                                                       & 0.9286                                                            & 0.9615                                                              & 0.9615                                                               & \textbf{0.9630}                                                        \\
\multicolumn{1}{c|}{\textbf{FS}}  & 0.9091                                                            & 0.9091                                                              & 0.9333                                                               & \textbf{0.9375}                                                       & 0.7273                                                            & 0.7000                                                              & 0.7826                                                               & \textbf{0.8182}                                                       & 0.8421                                                            & 0.8571                                                              & 0.8571                                                               & \textbf{0.9000}                                                          \\
\multicolumn{1}{c|}{\textbf{ACC}} & 0.9189                                                            & 0.9189                                                              & \textbf{0.9459}                                                      & \textbf{0.9459}                                                       & 0.8378                                                            & 0.8378                                                              & 0.8649                                                               & \textbf{0.8919}                                                       & 0.9189                                                            & 0.9189                                                              & 0.9189                                                               & \textbf{0.9459}                                                       \\ \hline
\end{tabular}}
\end{table*}

\begin{table*}[h!]
\caption{Backbone selection: Validation results from different multi-class approaches in terms of micro and macro-averages.}
\label{validation_averages}
\centering
\renewcommand{\arraystretch}{1.3}
\resizebox{13cm}{!}{
\begin{tabular}{ccccc|cccc}
\hline
                                  & \multicolumn{4}{c|}{\textbf{Micro-Average}}                                                                                                                                                                                                                                            & \multicolumn{4}{c}{\textbf{Macro-Average}}                                                                                                                                                                                                                                             \\ \hline
                                  & \textit{\begin{tabular}[c]{@{}c@{}}RNFL \\ features \cite{garcia2020ideal}\end{tabular}} & \textit{\begin{tabular}[c]{@{}c@{}}Fine-tuned\\ VGG16 \cite{garcia2020icip}\end{tabular}} & \textit{\begin{tabular}[c]{@{}c@{}}Fine-tuned\\ RAGNet \cite{garcia2020CMPB}\end{tabular}} & \textit{\begin{tabular}[c]{@{}c@{}}OCT-hybrid\\ network\end{tabular}} & \textit{\begin{tabular}[c]{@{}c@{}}RNFL \\ features \cite{garcia2020ideal}\end{tabular}} & \textit{\begin{tabular}[c]{@{}c@{}}Fine-tuned\\ VGG16 \cite{garcia2020icip}\end{tabular}} & \textit{\begin{tabular}[c]{@{}c@{}}Fine-tuned\\ RAGNet \cite{garcia2020CMPB}\end{tabular}} & \textit{\begin{tabular}[c]{@{}c@{}}OCT-hybrid\\ network\end{tabular}} \\ \hline
\multicolumn{1}{c|}{\textbf{SN}}  & 0.8378                                                            & 0.8378                                                              & 0.8649                                                               & \textbf{0.8919}                                                       & 0.8222                                                            & 0.8278                                                              & 0.8611                                                               & \textbf{0.8833}                                                       \\
\multicolumn{1}{c|}{\textbf{SP}}  & 0.9189                                                            & 0.9189                                                              & 0.9324                                                               & \textbf{0.9459}                                                       & 0.9155                                                            & 0.9165                                                              & 0.9335                                                               & \textbf{0.9440}                                                       \\
\multicolumn{1}{c|}{\textbf{PPV}} & 0.8378                                                            & 0.8378                                                              & 0.8649                                                               & \textbf{0.8919}                                                       & 0.8407                                                            & 0.8422                                                              & 0.8566                                                               & \textbf{0.8941}                                                       \\
\multicolumn{1}{c|}{\textbf{NPV}} & 0.9189                                                            & 0.9189                                                              & 0.9324                                                               & \textbf{0.9459}                                                       & 0.9268                                                            & 0.9297                                                              & 0.9336                                                               & \textbf{0.9506}                                                       \\
\multicolumn{1}{c|}{\textbf{FS}}  & 0.8378                                                            & 0.8378                                                              & 0.8649                                                               & \textbf{0.8919}                                                       & 0.8262                                                            & 0.8221                                                              & 0.8577                                                               & \textbf{0.8852}                                                       \\
\multicolumn{1}{c|}{\textbf{ACC}} & 0.8919                                                            & 0.8919                                                              & 0.9099                                                               & \textbf{0.9279}                                                       & 0.8919                                                            & 0.8919                                                              & 0.9099                                                               & \textbf{0.9279}                                                       \\ \hline
\end{tabular}}
\end{table*}

\begin{table*}[h!]
\caption{Learning strategy: Validation results per class using the proposed OCT-hybrid backbone for glaucoma grading.}
\label{val_proto_perclass}
\centering
\renewcommand{\arraystretch}{1.3}
\resizebox{15cm}{!}{
\begin{tabular}{cccc|ccc|ccc}
\hline
                                  & \multicolumn{3}{c|}{\textbf{HEALTHY}}                                                                                                                                                                                      & \multicolumn{3}{c|}{\textbf{EARLY GLAUCOMA}}                                                                                                                                                                               & \multicolumn{3}{c}{\textbf{ADVANCED GLAUCOMA}}                                                                                                                                                                             \\ \hline
                                  & \textit{\begin{tabular}[c]{@{}c@{}}Conventional\\ multi-class\end{tabular}} & \textit{\begin{tabular}[c]{@{}c@{}}Static\\ prototypes\end{tabular}} & \textit{\begin{tabular}[c]{@{}c@{}}Dynamic\\ prototypes\end{tabular}} & \textit{\begin{tabular}[c]{@{}c@{}}Conventional\\ multi-class\end{tabular}} & \textit{\begin{tabular}[c]{@{}c@{}}Static\\ prototypes\end{tabular}} & \textit{\begin{tabular}[c]{@{}c@{}}Dynamic\\ prototypes\end{tabular}} & \textit{\begin{tabular}[c]{@{}c@{}}Conventional\\ multi-class\end{tabular}} & \textit{\begin{tabular}[c]{@{}c@{}}Static\\ prototypes\end{tabular}} & \textit{\begin{tabular}[c]{@{}c@{}}Dynamic\\ prototypes\end{tabular}} \\ \hline
\multicolumn{1}{c|}{\textbf{SN}}  & \textbf{1}                                                                  & \textbf{1}                                                           & \textbf{1}                                                            & 0.7500                                                                      & 0.7500                                                               & \textbf{0.8333}                                                       & \textbf{0.9000}                                                             & 0.8000                                                               & \textbf{0.9000}                                                          \\
\multicolumn{1}{c|}{\textbf{SP}}  & 0.9091                                                                      & 0.9091                                                               & \textbf{1}                                                            & \textbf{0.9600}                                                             & 0.9200                                                               & \textbf{0.9600}                                                       & \textbf{0.9630}                                                             & 0.9630                                                               & 0.9259                                                                \\
\multicolumn{1}{c|}{\textbf{PPV}} & 0.8824                                                                      & 0.8824                                                               & \textbf{1}                                                            & 0.9000                                                                      & 0.8182                                                               & \textbf{0.9091}                                                       & \textbf{0.9000}                                                             & 0.8889                                                               & 0.8182                                                                \\
\multicolumn{1}{c|}{\textbf{NPV}} & \textbf{1}                                                                  & \textbf{1}                                                           & \textbf{1}                                                            & 0.8889                                                                      & 0.8846                                                               & \textbf{0.9231}                                                       & \textbf{0.9630}                                                             & 0.9286                                                               & 0.9615                                                                \\
\multicolumn{1}{c|}{\textbf{FS}}  & 0.9375                                                                      & 0.9375                                                               & \textbf{1}                                                            & 0.8182                                                                      & 0.7826                                                               & \textbf{0.8696}                                                       & \textbf{0.9000}                                                             & 0.8421                                                               & 0.8571                                                                \\
\multicolumn{1}{c|}{\textbf{ACC}} & 0.9459                                                                      & 0.9459                                                               & \textbf{1}                                                            & 0.8919                                                                      & 0.8649                                                               & \textbf{0.9189}                                                       & \textbf{0.9459}                                                             & 0.9189                                                               & 0.9189                                                                \\ \hline
\end{tabular}}
\end{table*}

\begin{table*}[h!]
\caption{Learning strategy: Validation results in terms of micro and macro-averages using the proposed OCT-hybrid backbone.}
\label{val_proto_averages}
\centering
\renewcommand{\arraystretch}{1.3}
\resizebox{12cm}{!}{
\begin{tabular}{cccc|ccc}
\hline
                                  & \multicolumn{3}{c|}{\textbf{Micro-Average}}                                                                                                                                                                                & \multicolumn{3}{c}{\textbf{Macro-Average}}                                                                                                                                                                                 \\ \hline
                                  & \textit{\begin{tabular}[c]{@{}c@{}}Conventional\\ multi-class\end{tabular}} & \textit{\begin{tabular}[c]{@{}c@{}}Static\\ prototypes\end{tabular}} & \textit{\begin{tabular}[c]{@{}c@{}}Dynamic\\ prototypes\end{tabular}} & \textit{\begin{tabular}[c]{@{}c@{}}Conventional\\ multi-class\end{tabular}} & \textit{\begin{tabular}[c]{@{}c@{}}Static\\ prototypes\end{tabular}} & \textit{\begin{tabular}[c]{@{}c@{}}Dynamic\\ prototypes\end{tabular}} \\ \hline
\multicolumn{1}{c|}{\textbf{SN}}  & 0.8919                                                                      & 0.8649                                                               & \textbf{0.9189}                                                       & 0.8833                                                                      & 0.8500                                                               & \textbf{0.9111}                                                       \\
\multicolumn{1}{c|}{\textbf{SP}}  & 0.9459                                                                      & 0.9324                                                               & \textbf{0.9595}                                                       & 0.9440                                                                      & 0.9307                                                               & \textbf{0.9620}                                                       \\
\multicolumn{1}{c|}{\textbf{PPV}} & 0.8919                                                                      & 0.8649                                                               & \textbf{0.9189}                                                       & 0.8941                                                                      & 0.8631                                                               & \textbf{0.9091}                                                       \\
\multicolumn{1}{c|}{\textbf{NPV}} & 0.9459                                                                      & 0.9324                                                               & \textbf{0.9595}                                                       & 0.9506                                                                      & 0.9377                                                               & \textbf{0.9615}                                                       \\
\multicolumn{1}{c|}{\textbf{FS}}  & 0.8919                                                                      & 0.8649                                                               & \textbf{0.9189}                                                       & 0.8852                                                                      & 0.8541                                                               & \textbf{0.9089}                                                       \\
\multicolumn{1}{c|}{\textbf{ACC}} & 0.9279                                                                      & 0.9099                                                               & \textbf{0.9459}                                                       & 0.9279                                                                      & 0.9099                                                               & \textbf{0.9459}                                                       \\ \hline
\end{tabular}}
\end{table*}

\begin{table*}[h!]
\caption{Validation accuracy reached by the dynamic prototypical approach using different statistics  and distance metrics.}
\label{distance_table}
\centering
\renewcommand{\arraystretch}{1.3}
\resizebox{14cm}{!}{
\begin{tabular}{ccccccccc}
\hline
\textbf{Distance}  & Euclidean & Cosine & Manhattan & Canberra & \textbf{Euclidean} & Cosine & Manhattan & Canberra \\
\textbf{Statistic} & median    & median & median    & median   & \textbf{mean}      & mean   & mean      & mean     \\ \hline
\textbf{Accuracy}  & 0.9279    & 0.9279 & 0.9099    & 0.8559   & \textbf{0.9459}    & 0.9099 & 0.9279    & 0.8739   \\ \hline
\end{tabular}}
\end{table*}

%% file: 5_Prediction_results.tex
\section{Prediction results}\label{sec:05_Prediction_Results}

\subsection{Quantitative results}
\begin{figure}[t]
\begin{center}
\includegraphics[width=7.5cm]{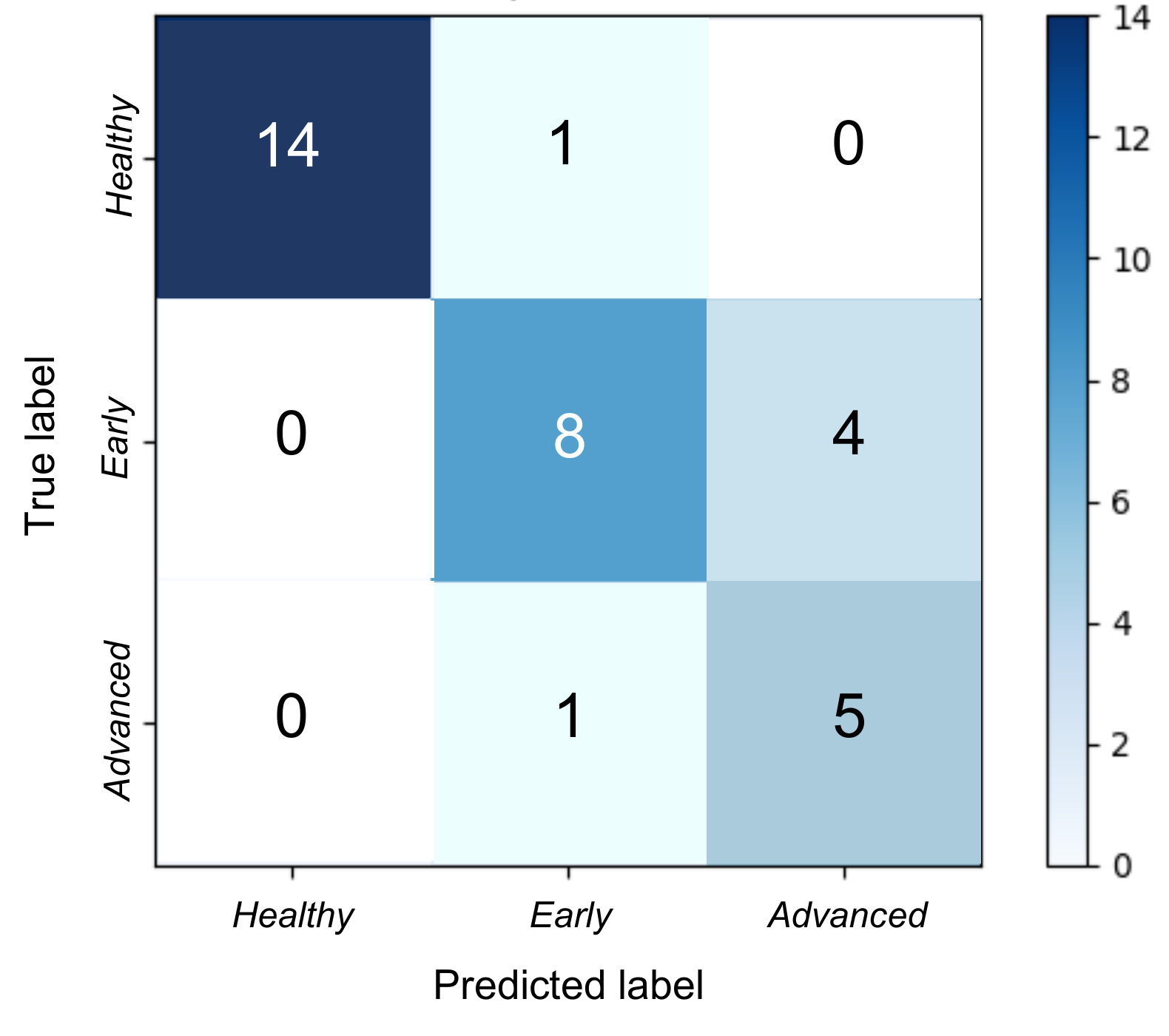}\\
\end{center}
\caption{Confusion matrix obtained by the dynamic prototype-based approach during the prediction phase.}
\label{confusion_matrix}
\end{figure}

In this section, we show the quantitative results achieved by the three learning strategies conducted during the prediction of the test set. It is worth noting that all the approaches contrasted here were addressed using the proposed OCT-based hybrid backbone as a feature extractor since it reported the best results in the validation phase. As before, we evaluate the models' performance both per class (Table \ref{test_results_perclass}) and in terms of average (Table \ref{test_results_average}), using different figures of merit. Additionally, in Fig. \ref{confusion_matrix}, we show the confusion matrix obtained by the best approach, i.e. the dynamic prototype-based model, to evidence the overall behaviour of the proposed method when predicting new samples. 

\begin{figure}[h!]
\begin{center}
\includegraphics[width=8cm]{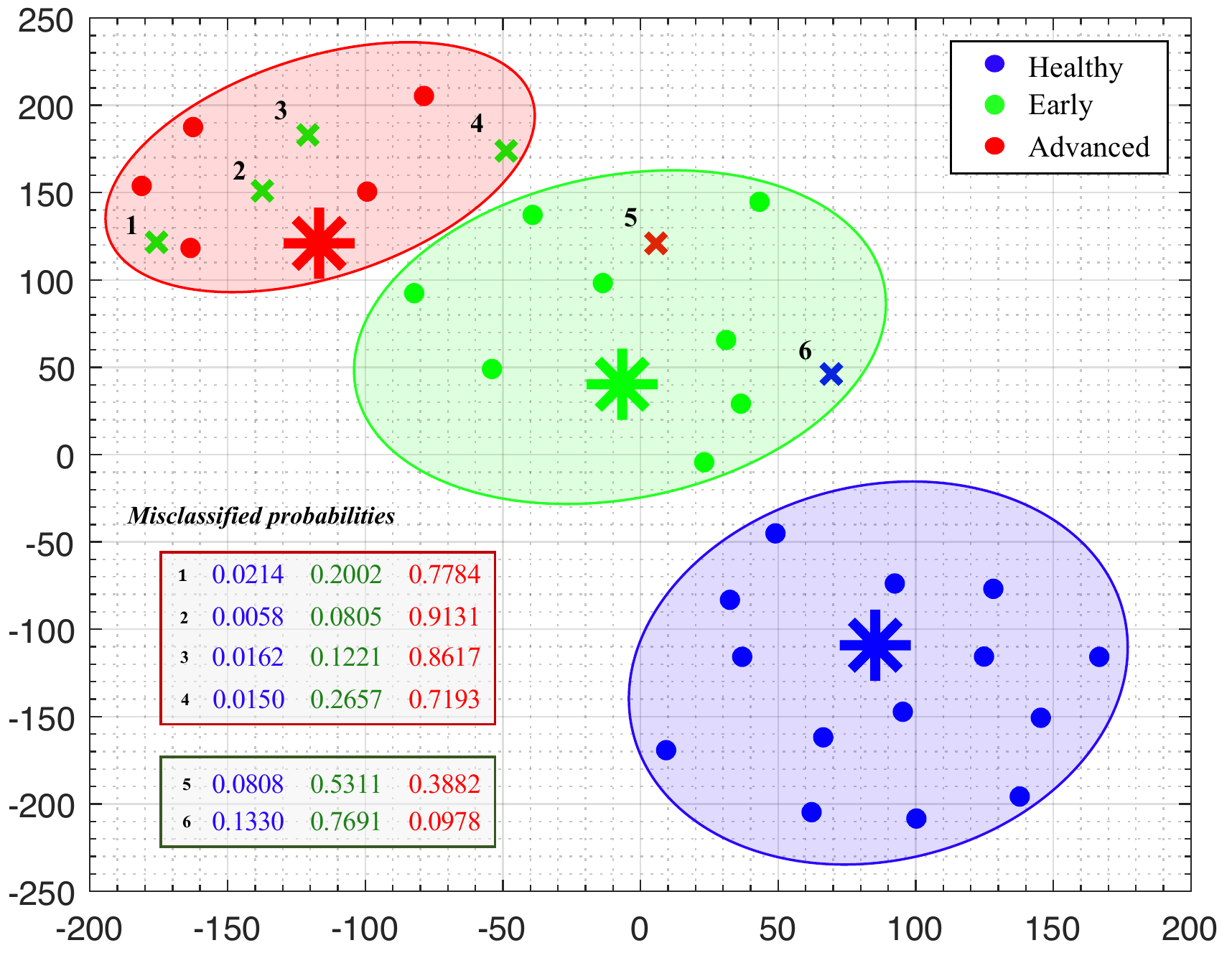}\\
\end{center}
\caption{Latent space showing the prototypes and the embedding test features on a 2D map via t-distributed Stochastic Neighbor Embedding (TSNE) tool. Blue, green and red colours denote the ground truth labels of the healthy, early and advanced representations, respectively.}
\label{TSNE}
\end{figure}

Also, to provide a more comprehensive interpretation of the glaucoma grading scenario, we illustrate in Fig. \ref{TSNE} a 2D map corresponding to the latent space arranged by the dynamic learning. In particular, the prototypes (denoted by asterisks) were calculated from the training and validation sets, whereas spots and crosses make reference to the embedding representations of the well and miss-classified test data, respectively. In addition to this, Euclidean distance-based probabilities reported from the miss-classified samples are also detailed in Fig. \ref{TSNE} to manifest the confidence of the dynamic model when it is wrong in the prediction.

\subsection{Qualitative results}
Class activation maps (CAMs) \cite{zhou2016} were computed to remark the regions in which the proposed dynamic prototypical network paid attention to predict the class of the test samples. The reported heatmaps allow a better understanding of the CNN-extracted features by highlighting the most relevant information of the B-scan for the predictions. These activated maps may provide an additional interpretation of the results for glaucoma grading depending on the patterns highlighted for each class. However, it should be mentioned that CAMs usually do not present a high precision at pixel-level since the heatmaps are created from the last layers of the proposed model, which output a down-sampled map from the input image.

\begin{figure*}[h!]
\begin{center}
\includegraphics[width=12.5cm]{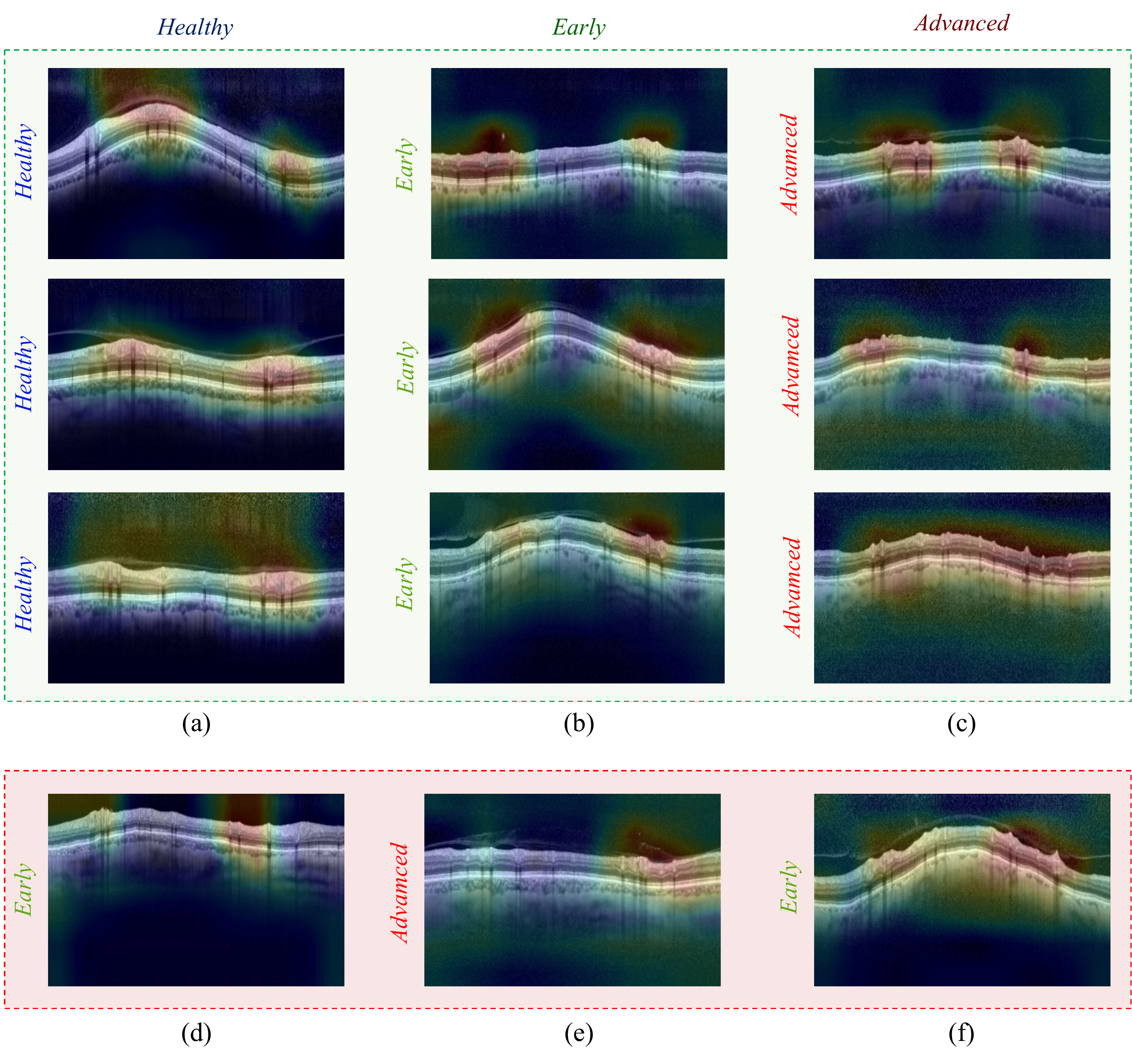}\\
\end{center}
\caption{CAMs showing the regions of interest in which the proposed prototypical-dynamic model pays attention to predict each class. Heatmaps over the green background (a-c) correspond to well-classified images, whereas the red frame (d-f) represents the miss-classified samples.}
\label{CAMs}
\end{figure*}

In Fig. \ref{CAMs}, we show several examples of CAMs corresponding to correctly and wrongly predicted samples to elucidate the relevant patterns found by the dynamic prototypical network to address the B-scans prediction. Specifically, we expose three well-classified heatmaps for each class to demonstrate the criteria followed by the proposed model to determine the predicted labels. Also, we report in the red frame of Fig. \ref{CAMs} some examples of miss-classified B-scans to visually evidence the reason why the model gets wrong in the prediction. Note that rows and columns in the illustration represent the predictions and labels, respectively.

\begin{table*}[h]
\caption{Prediction stage: Test results per class reached by the different proposed learning strategies for glaucoma grading.}
\label{test_results_perclass}
\centering
\renewcommand{\arraystretch}{1.1}
\resizebox{14cm}{!}{
\begin{tabular}{cccc|ccc|ccc}
\hline
                                  & \multicolumn{3}{c|}{\textbf{HEALTHY}}                                                                                                                                                                                      & \multicolumn{3}{c|}{\textbf{EARLY GLAUCOMA}}                                                                                                                                                                               & \multicolumn{3}{c}{\textbf{ADVANCED GLAUCOMA}}                                                                                                                                                                             \\ \hline
\textbf{}                         & \textit{\begin{tabular}[c]{@{}c@{}}Conventional\\ multi-class\end{tabular}} & \textit{\begin{tabular}[c]{@{}c@{}}Static\\ prototypes\end{tabular}} & \textit{\begin{tabular}[c]{@{}c@{}}Dynamic\\ prototypes\end{tabular}} & \textit{\begin{tabular}[c]{@{}c@{}}Conventional\\ multi-class\end{tabular}} & \textit{\begin{tabular}[c]{@{}c@{}}Static\\ prototypes\end{tabular}} & \textit{\begin{tabular}[c]{@{}c@{}}Dynamic\\ prototypes\end{tabular}} & \textit{\begin{tabular}[c]{@{}c@{}}Conventional\\ multi-class\end{tabular}} & \textit{\begin{tabular}[c]{@{}c@{}}Static\\ prototypes\end{tabular}} & \textit{\begin{tabular}[c]{@{}c@{}}Dynamic\\ prototypes\end{tabular}} \\ \hline
\multicolumn{1}{c|}{\textbf{SN}}  & \textbf{0.9333}                                                             & \textbf{0.9333}                                                      & \textbf{0.9333}                                                       & 0.5000                                                                      & 0.4167                                                               & \textbf{0.6667}                                                       & \textbf{1}                                                                  & 0.8333                                                               & 0.8333                                                                \\
\multicolumn{1}{c|}{\textbf{SP}}  & \textbf{1}                                                                  & \textbf{1}                                                           & \textbf{1}                                                            & \textbf{0.9524}                                                             & 0.9048                                                               & 0.9048                                                                & 0.7778                                                                      & 0.7407                                                               & \textbf{0.8519}                                                       \\
\multicolumn{1}{c|}{\textbf{PPV}} & \textbf{1}                                                                  & \textbf{1}                                                           & \textbf{1}                                                            & \textbf{0.8571}                                                             & 0.7143                                                               & 0.8000                                                                & 0.5000                                                                      & 0.4167                                                               & \textbf{0.5556}                                                       \\
\multicolumn{1}{c|}{\textbf{NPV}} & \textbf{0.9474}                                                             & \textbf{0.9474}                                                      & \textbf{0.9474}                                                       & 0.7692                                                                      & 0.7308                                                               & \textbf{0.8210}                                                       & \textbf{1}                                                                  & 0.9524                                                               & 0.9583                                                                \\
\multicolumn{1}{c|}{\textbf{FS}}  & \textbf{0.9655}                                                             & \textbf{0.9655}                                                      & \textbf{0.9655}                                                       & 0.6316                                                                      & 0.5263                                                               & \textbf{0.7273}                                                       & \textbf{0.6667}                                                             & 0.5556                                                               & \textbf{0.6667}                                                       \\
\multicolumn{1}{c|}{\textbf{ACC}} & \textbf{0.9697}                                                             & \textbf{0.9697}                                                      & \textbf{0.9697}                                                       & 0.7879                                                                      & 0.7273                                                               & \textbf{0.8182}                                                       & 0.8182                                                                      & 0.7576                                                               & \textbf{0.8485}                                                       \\ \hline
\end{tabular}}
\end{table*}

\begin{table*}[h]
\caption{Prediction stage: Test results reached in terms of micro and macro-average by the different proposed learning strategies.}
\label{test_results_average}
\centering
\renewcommand{\arraystretch}{1.1}
\resizebox{10cm}{!}{
\begin{tabular}{cccc|ccc}
\hline
                                  & \multicolumn{3}{c|}{\textbf{Micro-Average}}                                                                                                                                                                                & \multicolumn{3}{c}{\textbf{Macro-Average}}                                                                                                                                                                                 \\ \hline
                                  & \textit{\begin{tabular}[c]{@{}c@{}}Conventional\\ multi-class\end{tabular}} & \textit{\begin{tabular}[c]{@{}c@{}}Static\\ prototypes\end{tabular}} & \textit{\begin{tabular}[c]{@{}c@{}}Dynamic\\ prototypes\end{tabular}} & \textit{\begin{tabular}[c]{@{}c@{}}Conventional\\ multi-class\end{tabular}} & \textit{\begin{tabular}[c]{@{}c@{}}Static\\ prototypes\end{tabular}} & \textit{\begin{tabular}[c]{@{}c@{}}Dynamic\\ prototypes\end{tabular}} \\ \hline
\multicolumn{1}{c|}{\textbf{SN}}  & 0.7879                                                                      & 0.7273                                                               & \textbf{0.8182}                                                       & 0.8111                                                                      & 0.7278                                                               & \textbf{0.8112}                                                       \\
\multicolumn{1}{c|}{\textbf{SP}}  & 0.8939                                                                      & 0.8636                                                               & \textbf{0.9091}                                                       & 0.9101                                                                      & 0.8818                                                               & \textbf{0.9189}                                                       \\
\multicolumn{1}{c|}{\textbf{PPV}} & 0.7879                                                                      & 0.7273                                                               & \textbf{0.8182}                                                       & 0.7852                                                                      & 0.7103                                                               & \textbf{0.7857}                                                       \\
\multicolumn{1}{c|}{\textbf{NPV}} & 0.8939                                                                      & 0.8636                                                               & \textbf{0.9091}                                                       & 0.9055                                                                      & 0.8768                                                               & \textbf{0.9106}                                                       \\
\multicolumn{1}{c|}{\textbf{FS}}  & 0.7879                                                                      & 0.7273                                                               & \textbf{0.8182}                                                       & 0.7446                                                                      & 0.6825                                                               & \textbf{0.7865}                                                       \\
\multicolumn{1}{c|}{\textbf{ACC}} & 0.8586                                                                      & 0.8182                                                               & \textbf{0.8788}                                                       & 0.8586                                                                      & 0.8182                                                               & \textbf{0.8788}                                                       \\ \hline
\end{tabular}}
\end{table*}

%% file: 6_Discussion.tex
\section{Discussion}\label{sec:06_Discussion}


It should be noted that OCT B-scans contain information too limited to provide a complex diagnosis based on glaucoma grading. So, additional tests are usually carried out to accomplish a more exhaustive and reliable glaucoma grading diagnosis considering all the stages of the disease. Currently, the OCT systems can provide a result indicating if the B-scan is within normal limits, borderline or outside normal limits, which is equivalent to discern between healthy, suspect and glaucomatous cases. At this point, our system has demonstrated to provide a very high performance achieving an accuracy of $100\%$ and $96,97\%$ during the validation and testing phases, respectively. However, with the proposed system, we pretend to go beyond the binary classification of healthy vs glaucoma by means of a prototypical framework able to discern early from advanced glaucoma only using OCT samples, which adds significant value to the body of knowledge. Below, we discuss about the different CNN configurations and learning strategies carried out in this work for glaucoma grading.

\subsection{About the ablation experiments}

\textbf{Backbone selection}.
One of the main novelties addressed in this paper lies in the proposed OCT-based hybrid network, which combines hand-crafted and automatic features extracted from the input B-scans. In order to elucidate the superiority of the proposed model with respect to similar approaches, we raise a multi-class glaucoma-grading scenario comparing different methodologies. Note that a direct comparison with other state-of-the-art studies was not possible because of the lack of public repositories with annotations of different glaucoma severity levels, as claimed in \cite{raja2020}. For that reason, in this paper, we contrasted the proposed model with other own glaucoma-detection methods recently published in \cite{garcia2020icip, garcia2020ideal, garcia2020CMPB}. It should be mentioned that all the approaches were adapted to the multi-class environment to provide a reliable comparison during the validation stage, as detailed in Tables \ref{validation_perclass} and \ref{validation_averages}. Particularly, from Table \ref{validation_perclass} we can observe that the fine-tuned RAGNet model \cite{garcia2020CMPB} introduces slight improvements as compared with the traditional VGG16 architecture and the hand-crafted RNFL features. The use of tailored residual blocks and attention modules allows surpassing the results for most of the metrics by providing more distinctive feature maps. However, the proposed OCT-based hybrid network outperforms the rest of the models in the discrimination of all the classes. Focusing on the healthy column, RNFL features, VGG16 architecture and the proposed hybrid backbone report a more sensible behaviour highlighting for the SN and NPV metrics, whereas RAGNet model showcases more specific results by outstanding for the SP and PPV figures of merit. More global metrics, such as FS and ACC show higher values for RAGNet and OCT-hybrid approaches. Contrarily, the models differ to a greater extent in the discrimination of the glaucoma severity grades. Specifically, the proposed OCT-hybrid network report the best performance distinguishing between early and advanced glaucomatous samples for all the measures. The results detailed in Table \ref{validation_averages} further strengthens our confidence in the proposed OCT-hybrid network since it reports the higher values for all figures of merit in terms of micro and macro-average. Table \ref{validation_averages} is especially interesting to compare the model's performance since it gives an idea of their overall precision. In such a table, we can observe that hand-driven and basic deep learning approaches present a similar behaviour, whereas more sophisticated CNNs provide substantial performance improvements. Nevertheless, the combination of hand-crafted RNFL-based features and refined CNN architectures yields the best multi-class model for glaucoma grading, achieving an average accuracy of 0.9279.

\BlankLine
\textbf{Prototype-based learning strategy}.
The best approach reported during the validation phase was used as a feature extractor to address the next stage corresponding to the prototype-based learning strategy. So, according to the previous statements, the proposed OCT-hybrid network $\Psi_\phi$ was selected as the backbone of the prototypical architectures. At this point, several contributions are proposed in a novel framework for glaucoma grading. The method based on static prototypes was conducted for the first time by inferring the weights ($\phi \rightarrow \Theta$) of class-based prediction networks, instead of auto-encoders, as in the case of CBIR-based studies \cite{petscharnig2017, song2017, daoud2019}, or architectures intended to discern between positive and negative classes, as in the contrastive learning works \cite{khosla2020, chen2020, le2020}. Notwithstanding, the main novelties related to the prototypical environment were introduced in the dynamic approximation. As detailed in Section \ref{3_2_2_Dynamic_prototypes}, state-of-the-art studies usually made use of $\mathcal{K}\in[1,5]$ labelled images to extract the prototypes and $\mathcal{U}=$ 5, 10 or 15 unlabelled query samples to measure the latent distance to the prototypes during the training of the models. Unlike them, in this paper, the few-shot paradigm was redefined by dealing the $\mathcal{K}$ and $\mathcal{U}=P-\mathcal{K}$ variables as any other hyper-parameter to be optimized in the validation phase. As another novelty, we made use of labelled samples to build the query set $\mathcal{Q}$ in order to make the most of all available information about the B-scans for glaucoma grading. Our hypothesis claimed that the use of more than $\mathcal{K}=5$ support images to extract the prototypes would provide better results. However, it would be necessary to find the optimal balance between the number of support and query samples, because the data used to extract the prototypes $\rho_c$ is just as important as the data employed to know how well the prototypes work. For that reason, we show in Fig. \ref{val_shot} the categorical accuracy and the loss value achieved by the dynamic prototypical model using $\mathcal{K}\in[1,P]$ samples, being $P=41$ the total number of the training samples of the minority class (see Table \ref{partition}). As expected, the model's performance using a few $\mathcal{K}$ samples remains low and it improves as the number of support samples increases. However, the model reaches the best in the middle of the plotting, and then, the curve stabilizes or even worsens. This fact consolidates our hypothesis which holds the importance of a similar splitting of the support and query samples to train the models. In particular, the highest performance is reached using $\mathcal{K}=20$ shots and $\mathcal{U}=21$ query scans, which reports a validation accuracy of 0.9459 and a loss value of 0.3209, according to Fig. \ref{val_shot}. 

In Tables \ref{val_proto_perclass} and \ref{val_proto_averages}, both static and dynamic prototype-based approaches are compared with a conventional multi-class system based on the OCT-hybrid network trained for glaucoma grading. The classification of healthy samples in Table \ref{val_proto_perclass} deserves special attention since the dynamic prototypical model perfectly works for detecting the non-glaucomatous data, achieving the $100\%$ of performance for all figures of merit. The discrimination of the middle class, i.e. early class, makes the outperforming of the dynamic prototypical network evident. However, the conventional multi-class strategy exceeds the results for advanced glaucoma cases in all the metrics. Similar results can be appreciated in Table \ref{val_proto_averages}, which provides a comparison of the models' behaviour in terms of micro and macro-average. Specifically, the dynamic prototype reaches the best results with values higher than 0.90 for all the figures of merit. It should be highlighted that the baseline approach outperforms the strategy based on static prototypes, which reveals that end-to-end methodologies provide more robust models, as expected.

Furthermore, we carried out an empirical exploration of different distance metrics and statistical parameters to maximize the classification agreement in the latent space between the prototypes and the embedding query representations. Particularly, statistics based on mean and median operations were considered to extract the prototypes from the support samples, whereas Euclidean, Cosine, Manhattan and Canberra distances were subject to study to determine the class of the query representations according to the closest prototype in the latent space. As detailed in Table \ref{distance_table}, the highest validation accuracy was reached using the Euclidean distance and the mean statistic, which reported a validation accuracy of 0.9459.

\subsection{About the prediction results}

\textbf{Quantitative results}.
In this section, we compare the performance of the different proposed learning strategies during the prediction of the test set. As in the validation phase, results per class and in terms of micro and macro-average are reported in Tables \ref{test_results_perclass} and \ref{test_results_average}, respectively. According to the previous section, the discrimination between healthy and glaucoma classes is successfully accomplished by achieving results higher than 0.93 in the prediction of healthy samples for all figures of merit (see Table \ref{test_results_perclass}). It is remarkable that the three contrasted models provide the same effectiveness in distinguishing the healthy class. However, the glaucoma grading results make clear the differences between the models. In particular, the dynamic prototypical network shows better SN and NPV results, whereas the baseline approach out stands for SP and PPV metrics in the early glaucoma detection. The opposite happens in the advanced glaucoma case since the conventional approach reveals a more sensitive behaviour (higher recall and NPV) and the dynamic prototypical model provides more specific results (better specificity and precision). Note that F-score (FS) and Accuracy (ACC) metrics reach higher values using the dynamic prototypical strategy. Additionally, the results in Table \ref{test_results_average} are directly in line with those reported in the validation phase, since the dynamic prototypical method is consolidated as the best model, followed by the conventional multi-class approach and leaving the static prototypical network in the last position. From here, we can conclude that end-to-end trained systems are more compelling and allow providing more reliable and robust models. 

From the confusion matrix in Fig. \ref{confusion_matrix}, promising results are evidenced for glaucoma detection since the dynamic prototype strategy almost perfectly distinguishes glaucomatous from healthy samples. Only a specific healthy B-scan was wrongly predicted as a sample with early glaucoma. However, the proposed model makes more mistakes predicting the early class since it sometimes confuses early and advanced patterns of the disease. Nonetheless, in the case of the advanced glaucoma class, the prototypical network only disagrees with the expert ophthalmologist once, in which a severe glaucomatous scan was classified as early glaucoma.

In order to visualise the results of the confusion matrix in a more interpretable scenario, we illustrated the latent space environment arranged by the proposed dynamic prototypical network in Fig. \ref{TSNE}. The TSNE technique allows showing the distribution of the embedding test representations on a 2D map, as well as the prototypes extracted from the training samples. As appreciated in Fig. \ref{TSNE}, three well-differentiated sub-spaces arise on the 2D map locating the healthy and advanced representations in the opposite corners and the prototypes of early and advanced classes very close, as expected. Additionally, early glaucoma-related features are mapped in the middle of the plotting, which matches with the order of the glaucoma severity scale. In addition, the prediction probabilities for the miss-classified samples are detailed in Fig. \ref{TSNE} to demonstrate the coherence of the proposed model using the embedding features for glaucoma grading. For example, the nicknamed representations $1-4$ correspond to early glaucomatous samples which were miss-classified as advanced glaucoma items with a level of confidence less than 0.8 in some cases. Regarding the tagged representation $5$, the dynamic prototypical network wrongly predicted an early B-scan as an advanced sample, but with serious difficulties since the prediction probabilities are 0.53 and 0.39 for early and advanced classes, respectively. Something similar happens for the nicknamed point $6$, which was miss-classified as an early glaucoma sample when it was actually healthy. In spite of this, the embedding representation $6$ is located near the decision boundary between early and healthy classes, which manifests the robustness of the model despite being wrong. 

\BlankLine
\textbf{Qualitative results.}
The findings from the Class Activation Maps (CAMs) reported in Fig. \ref{CAMs} keep consistency with the clinical interpretation provided by the ophthalmologists, who claim that a thinning of the RNFL structure is usually associated with glaucomatous patterns, whereas a thickening of the RNFL denotes cues of healthy samples \cite{medeiros2012, leung2012}. As appreciated in Fig. \ref{CAMs}, the proposed model pays special attention to specific regions of the RNFL depending on the reported prediction, according to the outcomes found in our previous work \cite{garcia2020icip}. 

In the green frame of Fig. \ref{CAMs}, three random examples are illustrated to demonstrate that there is a repeating pattern for each of the classes. Well-classified samples corresponding to the healthy class (Fig. \ref{CAMs}. (a)) show heatmaps with highlighted pixels in regions characterised by a thickening of the RNFL. Oppositely, well-classified early and advanced glaucomatous samples (Fig. \ref{CAMs}. (b), (c)) provide apparent cues of RNFL deterioration, which is visible in the heatmaps highlighting regions where a thinning of the RNFL is evident.
As observed, the degradation level of the RNFL thickness is accentuated the greater the severity of the disease.

The wrongly predicted CAMs can be also appreciated in the red frame of Fig. \ref{CAMs}. Particularly, the example showed in Fig. \ref{CAMs} (d) shows a healthy B-scan miss-classified as an early glaucoma sample. Nevertheless, the model's decision shows to be coherent with the established patterns, since the B-scan presents several areas in which the RNFL appears slightly thinned. Something similar happens with the remainder of cases of the red frame. Specifically, the example in Fig. \ref{CAMs} (e) is striking since the RNFL seems to be widely deteriorated throughout the entire B-scan, so the model associates the image with an advanced glaucomatous sample. The opposite occurs in Fig. \ref{CAMs} (f), where an advanced glaucoma OCT image is predicted with the early label. In this case, the B-scan shows several regions in which the RNFL thickness agrees more with early than advanced glaucoma patterns. Therefore, from the red frame of Fig. \ref{CAMs}, it is possible to elucidate the complexity in the diagnosis of the different grades of glaucoma, even for expert ophthalmologists, who often disagree. It is important to remember at this point that, in this paper, we only made use of raw circumpapillary OCT images to build the predictive models. So, in future research lines, it would be interesting to include additional data in order to provide the model with the necessary information that clinicians take into account to determine the severity of the glaucoma disease.

%% file: 7_Conclusion.tex
\section{Conclusion}\label{sec:07_Conclusion}

In this paper, we have proposed several artificial intelligence solutions for glaucoma grading using raw circumpapillay OCT images. A new base encoder network has been developed improving the multi-class methodologies addressed to date for glaucoma severity detection. The proposed model introduces a residual convolutional block with tailored kernel sizes to get the most out from the B-scans and an attention module able to refine the features of the latent space to maximize the classification agreement. Besides, the encoder network uses, for the first time, a combination of hand-crafted and automatic features giving rise to an OCT-based hybrid model which adds substantial improvements in the final prediction. As a novelty, this architecture was employed as the backbone of a novel framework based on prototypical learning, in which the limits of the few-shot paradigm have been redefined for an optimal glaucoma grading procedure. This innovative approach carried out in an end-to-end manner has surpassed the multi-class baseline and it has reported promising results for both glaucoma detection and glaucoma grading scenarios, achieving testing accuracy of 0.9697 and 0.8788, respectively. In future research lines, the efforts should be focused on improving the discrimination of the different grades of glaucoma severity by including additional information outside the SD-OCT images. 